\documentclass[aps,prd,twocolumn, preprintnumbers, floatfix, superscriptaddress,showpacs]{revtex4}
\usepackage{graphicx}
\usepackage{amsmath,amsfonts,bm,bbold}
\usepackage{amssymb}
\begin{document}

\preprint{JLAB-THY-08-802}

\author{H.~R.~Grigoryan}
\affiliation{Thomas Jefferson National Accelerator Facility,
             Newport News, VA 23606, USA}
\affiliation{Physics Department, Louisiana State University,
             Baton Rouge, LA 70803, USA}
\author{A.~V.~Radyushkin}
\affiliation{Thomas Jefferson National Accelerator Facility,
              Newport News, VA 23606, USA}
\affiliation{Physics Department, Old Dominion University, Norfolk,
             VA 23529, USA}
\affiliation{Laboratory of Theoretical Physics, JINR, Dubna, Russian
             Federation}

\title{Anomalous Form Factor of the Neutral Pion in Extended AdS/QCD Model\\ with Chern-Simons Term}

\begin{abstract}
We propose an  extension of the  hard-wall AdS/QCD model  by  including  the Chern-Simons term required to
reproduce the chiral anomaly of QCD. In the framework of this 
holographic model,  we
study the vertex function $F_{\pi \gamma^* \gamma^*}(Q_1^2,Q_2^2) $ which accumulates 
information about  the coupling of the pion 
to two (in general virtual) 
photons. We calculate the slope 
of the  form  factor with one real and one
slightly virtual photon   and show that it is close to
experimental findings. 
We analyze the formal limit of  large virtualities and establish that 
predictions of the holographic model analytically 
(including nontrivial dependence on the ratio of photon virtualities) 
coincide  with those 
of  perturbative QCD calculated for the  asymptotic form of the pion distribution amplitude. 
We also investigate the generalized VMD structure 
of $F_{\pi \gamma^* \gamma^*}(Q_1^2,Q_2^2) $  in  the extended AdS/QCD
model.  
\end{abstract}


\maketitle

\section{Introduction}

The form factor $F_{\gamma^* \gamma^*  \pi^0}(Q_1^2,Q_2^2)$  describing the coupling of two (in general, virtual)
photons with the lightest hadron, the pion, plays a  special role in  the studies of  exclusive processes in
quantum chromodynamics (QCD). When both photons are real, the form factor $F_{\gamma^* \gamma^*  \pi^0}(0,0)$
determines the rate of the $\pi^0 \to \gamma \gamma$ decay, and its value at this point is deeply related to the
axial anomaly \cite{anomaly}. Because of this relation,   the $\gamma^* \gamma^*  \pi^0$ form factor was an object
of intensive studies since the 60's \cite{cornwall}-\cite{Jacob:1989pw}. 

At  large photon virtualities, its
behavior was studied \cite{Lepage:1980fj,Brodsky:1981rp,Lepage:1982gd} within perturbative QCD (pQCD)
factorization approach for exclusive processes \cite{Chernyak:1983ej,Efremov:1979qk,Lepage:1980fj}. Since  only
one hadron is involved, the
 $\gamma^* \gamma^*  \pi^0$ form factor has
the simplest  structure for pQCD analysis,  with
the nonperturbative information about the pion
accumulated in the pion distribution
amplitude $\varphi_\pi (x)$ introduced in Refs.~\cite{Radyushkin:1977gp,Lepage:1979zb}.
Another  simplification is that  the short-distance
amplitude for $\gamma^* \gamma^*  \pi^0$ vertex is given, 
at the leading order, just  by a single quark propagator.
Theoretically, most clean situation is when both photon virtualities
are large,  but  the  experimental study of
$F_{\gamma^* \gamma^*  \pi^0}(Q_1^2,Q_2^2)$  in this regime
through the  $\gamma^* \gamma^* \to \pi^0$ process
is very difficult due to very
small cross section.

The leading-twist pQCD factorization, however, 
works even if  one of the photons is real or almost
real.
Furthermore, this  kinematics is   amenable   to experimental investigation  through
the  $\gamma \gamma^* \to \pi^0$ process at $e^+e^-$ colliders.
Comparison of   the data obtained by   CELLO  \cite{Behrend:1990sr}
and CLEO \cite{Gronberg:1997fj}   collaborations
with the original  leading
\cite{Lepage:1980fj,Brodsky:1981rp} and next-to-leading order
\cite{delAguila:1981nk,Braaten:1982yp,Kadantseva:1985kb,Musatov:1997pu}
 pQCD predictions
amended by  later  studies \cite{Khodjamirian:1997tk,Schmedding:1999ap}
that incorporate
a more thorough treatment of the real photon channel
using the
light-cone QCD sum rule ideas
provided  important information about the shape
of the pion distribution amplitude $\varphi_{\pi}(x)$
(for the most recent review    see  \cite{Bakulev:2007jv}).
The
 momentum dependence of
$F_{\gamma^* \gamma^*  \pi^0}(0,Q^2)$ form factor was also studied in   various models of   nonperturbative QCD
dynamics (see, e.g.,
\cite{Voloshin:1982ea}-\cite{Ruiz
Arriola:2006ii} and references  therein).

In the present paper, our goal is to
extend the holographic dual model of QCD to incorporate the anomalous $F_{\gamma^* \gamma^*  \pi^0}(Q_1^2,Q_2^2)$
form factor.
During the last few years applications of gauge/gravity duality \cite{Maldacena:1997re} to hadronic physics
attracted a lot of attention, and  various holographic dual models of QCD were proposed in the literature (see,
e.g.,
\cite{Polchinski:2002jw}-\cite{Dhar:2007bz}).
These models were able to incorporate such essential properties of QCD as confinement and dynamical chiral
symmetry breaking, and also to reproduce many of the static hadronic observables (decay constants, masses), with
values rather  close to the experimental  ones.

As the basis for our extension,
we  follow the holographic approach of  Refs.~\cite{Erlich:2005qh,DaRold:2005zs}, and then 
intend to proceed
along the lines of formalism outlined in our recent papers \cite{Grigoryan:2007vg,Grigoryan:2007wn}, where it was
first applied to form factors and wave functions of vector mesons \cite{Grigoryan:2007vg,Grigoryan:2007iy}
(tensor   form factors of vector  mesons were considered in  \cite{Abidin:2008ku})  and later 
\cite{Grigoryan:2007wn}  to  the pion electromagnetic form factor.  However,  a
straightforward application of  the  approach of
Refs.~\cite{Erlich:2005qh,DaRold:2005zs,Grigoryan:2007vg,Grigoryan:2007wn}  to $F_{\gamma^* \gamma^*
\pi^0}(Q_1^2,Q_2^2)$ form factor   gives a vanishing result. There are two 
obvious reasons for such an outcome.

First, the five-dimensional (5D) gauge fields $B^a(x,z)$  in the AdS/QCD lagrangian of Refs.~\cite{Erlich:2005qh,DaRold:2005zs} are
only dual to the 4D isovector currents $J^a(x)$. On the other hand, a  nonzero result for the matrix element
$\langle 0| J_{\rm EM}^\mu  J_{\rm EM}^\nu |\pi^0 \rangle $ defining the $ F_{\gamma^* \gamma^*
\pi^0}(Q_1^2,Q_2^2)$ form factor  may be  obtained only when the electromagnetic currents  $J_{\rm EM}^\mu $, $
J_{\rm EM}^\nu$ have both isovector and isoscalar components,
 which is the case in two-flavor QCD, but not in the
holographic models of Refs.~\cite{Erlich:2005qh,DaRold:2005zs}.

Thus, we need an  AdS/QCD model that  includes  gauge fields in the 5D bulk which are dual to both
isovector and isoscalar currents. The natural way to do this is to extend the gauge group $SU(2)_L \otimes SU(2)_R
$ in the bulk up to $U(2)_L \otimes U(2)_R $.  After explicit separation of 
 the isosinglet  and isovector  parts, the new
5D field can be written as ${\cal B}_\mu =t^a \,B^a_\mu    +  \mathbb{1} \, \frac{ {\hat B}_\mu}{2} $, 
with 
 $\mathbb{1}$  being  the unity matrix.
The $ \hat  B $  part  is dual to 4D isosinglet vector current.

Even after this modification, the AdS/QCD action gives zero result for the correlator $\langle 0| J_\mu^{\{
I=1\} } J_\nu^{\{ I=0\} } J_\alpha^{A}|0 \rangle$ involving two  vector currents  $J_\mu^{\{ I=1\} }$, $ J_\nu^{\{
I=0\} }$  and the   axial current  $J_\alpha^{A}$ (that has nonzero projection onto  the pion state). To
bring in the anomalous amplitude into the model, the next step (similar to \cite{Domokos:2007kt}) is to add  
a Chern-Simons term \cite{Chern:1974ft}
to the  action. After these extensions, the calculation of the $ F_{\gamma^* \gamma^*
\pi^0}(Q_1^2,Q_2^2)$ form factor may be performed using the methods developed in
Refs.~\cite{Grigoryan:2007vg,Grigoryan:2007wn}.

The paper is organized in the following way. We start by recalling, in Section II, the basics of the hard-wall
model, in particular, the form of the  action given in Ref.~\cite{Erlich:2005qh} and
our results \cite{Grigoryan:2007wn} for the pion wave function. In  Section III, we consider the generalization of
the AdS/QCD model that includes isoscalar fields and Chern-Simons term. Using this extended model we describe the
calculation of the  $ F_{\gamma^* \gamma^* \pi^0}(Q_1^2,Q_2^2)$ form factor  and express it in terms of the pion
wave  function and two bulk-to-boundary propagators for the vector currents describing EM sources. In Section IV,
we study the results obtained within the extended AdS/QCD model with one real and one slightly virtual photon and
calculate  the value of the $Q^2$-slope of the form factor. We also discuss the formal limit of large photon
virtualities, and compare these results to those obtained in pQCD. In Section V, we study  the generalized VMD
structure of the AdS/QCD model expression for the $ F_{\gamma^* \gamma^* \pi^0}(Q_1^2,Q_2^2)$ form factor.
Finally, we summarize the paper.  

\section{Basics of Hard-Wall Model}

In the holographic model, QCD resonances correspond to Kaluza-Klein type  excitations of the sliced AdS$_5$ space
with the metric 
\begin{align}
g_{MN}dx^Mdx^N = \frac{1}{z^2}\left(\eta_{\mu \nu}dx^{\mu}dx^{\nu} -
dz^2\right) \ ,
\end{align}
where $ \eta_{\mu\nu} = {\rm Diag}\, (1,-1,-1,-1) $ and $ \mu, \nu = (0,1,2,3) $, $ M, N = (0,1,2,3,z) $.  
The basic  prescription is that there is a correspondence between the 4D vector and axial-vector currents and 5D
gauge fields $V^a_{\mu}(x,z)$ and $A^a_{\mu}(x,z)$. Furthermore, since the gauge invariance corresponding to the
axial-vector current is spontaneously broken in the 5D background, the longitudinal component of the axial-vector  field
becomes physical and related to the pion field.

\subsection{AdS/QCD action}

The action of the holographic model of Ref.~\cite{Erlich:2005qh} can be written in the form
\begin{widetext}
\begin{align}
\label{AdS} S^{B}_{\rm AdS} &= {\rm Tr}  \int d^4x \int_0^{z_0}
 dz~\left[\frac{1}{z^3}(D^{M}X)^{\dagger}(D_{M}X) +
\frac{3}{z^5} X^{\dagger}X - \frac{1}{8g_5^2z}(B_{(L)}^{MN}B_{(L)MN}+B_{(R)}^{MN}B_{(R) MN})\right] \  ,
\end{align}
\end{widetext}
where $ D X = \partial X - iB_{(L)}X + iX B_{(R)} $, ($ B_{(L,R)} = V \pm A $) and $X(x,z) = v(z)U(x,z)/2 $ is
taken as a product of the chiral field $ U(x,z) =  \exp{\left [2i t^a \pi^a(x,z)\right ]} $ (as usual,
$t^a=\sigma^a/2$, with $\sigma^a$   being Pauli  matrices) and the function 
$ v(z) = (m_q z + \sigma z^3) $
containing the chiral symmetry breaking parameters $m_q$ and $\sigma$, with  $m_q$ playing the role of  the quark
mass and $\sigma$  that of the  quark   condensate.

In general, one can write $ A = A_{\perp} + A_{\parallel} $, where $ A_{\perp} $ and $ A_{\parallel} $ are
transverse and longitudinal components of the axial-vector field. The spontaneous symmetry breaking causes $
A_{\parallel} $ to be physical and associated with the Goldstone boson, pion in this  case. The longitudinal
component is  written in the form:
\begin{align}
 A^a_{\parallel \, M}(x,z) = \partial_M \psi^a(x,z) \ .
\end{align}
Then $ \psi^a(x,z) $ corresponds to the pion field. This Higgs-like mechanism breaks the axial-vector gauge
invariance by bringing a $z$-dependent mass term in the $A$-part of the lagrangian.

Varying the action with respect to the  transverse gauge fields
$ V^a_{\mu}$ and
$A^a_{\perp \mu}$ gives equations of motion
for these fields describing ({\it via} the holographic correspondence)
the physics of vector and axial-vector   mesons.
Variation with respect to
 $A_{\parallel \mu}^a $ and
$A^a_{z} $  gives two coupled
equations for the chiral field $\pi^a (x,z)$ and the pion field
$\psi^a (x,z)$.
It is convenient to work in   Fourier
representation, where $ \tilde{V}^a_{\mu}(p,z) = \tilde{V}^a_{\mu}(p){\cal V} (p,z) $
 is the  Fourier transform of $
V^a_{\mu}(x,z) $, and $ \tilde{A}^a_{\mu}(p,z) $ is the Fourier
transform   of $ A^a_{\mu}(x,z) $.

\subsection{Vector  channel}

The vector bulk-to-boundary propagator
\begin{equation}
\label{Jmeson}  {\cal V } (p,z) = g_5\sum_{m = 1}^{\infty}\frac{ f_{m} \psi_m^V( z)}{ -p^2 + M^2_{m} }
\end{equation}
has the bound-state poles for $p^2=M_n^2$, with the resonance masses
$M_{n} = \gamma_{0,n}/z_0 $  ($ \gamma_{0,n}$ is $n^{\rm th}$ zero
of the Bessel function $J_0(x)$)  determined by the eigenvalues
of the equation of motion
\begin{align}
\partial_z\left[\frac{1}{z}\partial_z \psi^V_n(z)\right] + \frac{1}{z}M^2_n\psi^V_n(z) = 0 \ ,
\end{align}
subject to boundary conditions $ \psi^V_n(0) = \partial_z \psi^V_n(z_0) =0 $. 
The eigenfunctions of this equation 
give the ``$\psi$'' wave functions
\begin{equation}\label{holographicwf}
\psi_n^V ( z)   = \frac{\sqrt{2} }{z_0 J_1(\gamma_{0,n})}\, z J_1(M_{n}
z)
\end{equation}
for the relevant resonances.
The coupling constants $f_n$ are determined from  the $\psi$
wave  functions   through
\begin{equation}
 f_n =\frac1{g_5} \,  \left  [ \frac1{z}\, \partial_z \psi^V_n (z) \right ]_{z=0} =
\frac{\sqrt{2}  M_n} {g_5z_0  J_1 (\gamma_{0,n})} \ .
\end{equation}
In  Ref.~\cite{Grigoryan:2007vg}, we introduced  ``$\phi$ wave functions''
\begin{equation}
\label{phi}
 \phi_n^V (z) \equiv  \frac1{M_n z}\, \partial_z \psi_n^V (z) = \frac{\sqrt{2} }{z_0  J_1(\gamma_{0,n})}\,
J_0(M_{n} z)  \  ,
\end{equation}
which give   the couplings $g_5 f_n/M_n$ as   their values at   the origin,  just like the  ($L=0$) bound   state
wave functions  in  quantum mechanics. Moreover, these functions satisfy Dirichlet b.c. $\phi_n^V (z_0)=0$.
Physically, the $ \psi^V $ wave functions describe the vector  bound states in terms of the vector potential $
V_\mu $, while the $ \phi^V $ wave functions describe them in terms of the field-strength tensor $ \partial_z
V_\mu = V_{z\mu} $ (we work in the axial gauge, where $ V_z = 0 $).

An  essential ingredient of form   factor formulas is  the vector bulk-to-boundary propagator ${\cal J} (Q,z)
\equiv {\cal V}  (iQ,z)$  taken at a spacelike momentum $p$ with $p^2 =-Q^2$.  It  can be written in a closed form
as
\begin{equation}
\label{JQz} {\cal J} (Q,z) = {Qz}\left[K_1(Qz) + I_1(Qz) \frac{K_0(Qz_0)}{I_0(Qz_0)} \right] \, \ .
\end{equation}
The function ${\cal J} (Q,z)$ satisfies  the relations ${\cal J}(Q,0) = 1 $,  ${\cal J}(0,z) = 1$
and $ \partial_z{\cal J}(Q,z_0) = 0 $.

\subsection{Pion channel}

An important achievement of the hard-wall model of Ref.~\cite{Erlich:2005qh}  is its compliance with the
Gell-Mann--Oakes--Renner relation $m_\pi^2 \sim m_q$ that produces a massless pion in the $m_q=0$ limit. The fits
of Ref.~\cite{Erlich:2005qh} give very small value $m_q \sim 2$ MeV for the ``quark mass'' parameter $m_q$, so it
makes sense to resort to the  chiral $m_q=0$ limit, which has an additional advantage that solutions of equations of
motion in this case can be found analytically. The pion wave function  $\psi (z)$ is introduced through the
longitudinal part of the axial-vector field:
\begin{align}
\tilde{A}^a_{\parallel \mu}(p,z) = \frac{p_{\mu}p^{\alpha}}{p^2}\tilde{A}^a_{\alpha}(p)\psi(z) \ ,
\end{align}
The bulk-to-boundary part $\pi (z)$ of the  $\pi^a (z)$ field, in the zeroth order of the of $m_\pi^2$ expansion proposed
in Ref~\cite{Erlich:2005qh}  tends to $-1 $. Then the equation for $\Psi (z) \equiv \psi (z) -\pi(z)$ 
is exactly solvable, with the result
\begin{widetext}
\begin{align}
\label{Psi}
\Psi (z) = 
{z\, \Gamma \left ({2}/{3} \right  )
\left(\frac{\alpha}{2}\right)^{1/3}}
\left[ I_{-1/3}\left(\alpha z^3\right)  -  I_{1/3}\left(\alpha
z^3\right) \frac{I_{2/3}\left(\alpha z^3_0\right)} {I_{-2/3}\left(\alpha z^3_0\right)}\right]
 \ ,
\end{align}
\end{widetext}
where $ \alpha = g_5 \sigma/3$
(one may use here Airy  functions instead of $I_{-1/3}(x), I_{2/3} (x)$,
cf. \cite{Kwee:2007dd}).
The pion wave function $\Psi (z)$ coincides with the axial-vector bulk-to-boundary
propagator ${\cal A} (0,z)$ taken at $p^2=0$. It is normalized by $\Psi (0)=1$, satisfies Neumann boundary
condition $\Psi '(z_0)= 0$ at the IR boundary and, due to the holographic correspondence, has the property that
\begin{align}\label{piondecay2}
f^2_{\pi} =  - \frac{1}{g^2_5}\left(\frac{1}{z}\partial_z \Psi(z) \right)_{z = \epsilon \rightarrow 0}  \ .
\end{align}

For the neutral pion, it is  convenient to define $f_\pi$ through  the matrix element of the $\sigma_3$ projection
of the axial-vector  current
\begin{align}
 \langle 0 |  J_{\mu}^{A,3}  | \pi^0 (p) \rangle \equiv
\left  \langle 0 \left|  \frac{\bar u \gamma_\mu  \gamma_5  u - \bar d
 \gamma_\mu \gamma_5 d}{2} \right | \pi^0 (p) \right \rangle = i f_\pi p_\mu \  .
\end{align}
Then, the  (experimental) numerical value of $f_\pi$ is \mbox{$92.4$ MeV.} Matching  AdS$_5$ result $\Sigma^{\rm AdS}
(p^2)\sim \ln p^2/(2g_5^2)$ and QCD result $\Sigma_3^{\rm QCD} (p^2)\sim  \ln p^2/(8 \pi^2) $ for the
\mbox{large-$p^2$}
behavior of the correlator     of $J_\mu^{A, 3}$ currents gives \mbox{$g_5 = 2 \pi$ \cite{Erlich:2005qh}.}  Analyzing pion
EM form factor, one deals with  charged pions, and  the choice of the axial-vector 
 current as $J_{\mu}^{A,{\rm (c)} }=
\bar d \gamma_\mu  \gamma_5 u$ is  more   natural. Then $f_\pi^{\rm (c)} = 130.7$ MeV, while $\Sigma_{\rm (c)}^{\rm
QCD} (p^2)\sim  \ln p^2 /(4 \pi^2) $, hence, $g_5^{\rm (c)} = \sqrt{2} \pi$ \cite{Grigoryan:2007my}. Of course, the
combination $g_5^2 f_\pi^2$, being the ratio of the coefficient  of the pion pole contribution $\Sigma_\pi  (p^2) \sim
f_\pi^2/p^2$ to the coefficient  $\sim 1/g_5^2$  in  $\Sigma (p^2)$'s   \mbox{large-$p^2$}  behavior, remains intact:
\begin{align}
g_5^2 f_\pi^2 = 4 \pi^2 f_\pi^2 = 2  \pi^2 (f_\pi^{(c)})^2 =(g_5^{\rm (c)} f_\pi^{\rm (c)})^2 \equiv s_0/2 \  ,
\end{align}
where $s_0  \approx 0.67$ GeV$^2$. It is this convention-independent combination that enters
Eq.~(\ref{piondecay2}).

Again, it is convenient to   introduce the conjugate wave function \cite{Grigoryan:2007wn}:
\begin{align}
\label{Phi}
\Phi(z) = - \frac1{g_5^2 f_\pi^2} \left ( \frac1{z} \, \partial_z \Psi (z) \right)  =
- \frac{2}{s_0}\,  \left ( \frac1{z} \, \partial_z \Psi (z) \right)  \  .
\end{align}
It vanishes at the IR boundary $z=z_0$, i.e., $\Phi (z_0)=0$  and,  according to  Eq.~(\ref{piondecay2}), is  normalized as
$
 \Phi(0) = 1
$
at the origin.
From $\Phi (0) =1$, it follows that
  the pion decay constant can be written as  a function
\begin{align}
\label{fpi2}
{g_5^2}  f_\pi^2 = 3\cdot 2^{1/3} \, \frac{\Gamma (2/3) }{\Gamma (1/3) }\,
\frac{I_{2/3}\left(\alpha z^3_0\right)} {I_{-2/3}\left(\alpha
z^3_0\right)} \,  {\alpha^{2/3}}
\end{align}
of the  condensate parameter $\alpha$ and the confinement radius $z_0$. Note that the magnitude of $\alpha$ is
independent of the $g_5$-convention, while the value of $\sigma$  depends on the $g_5$-convention used.

After fixing $z_0$ through  the $\rho$-meson mass,  \mbox{$ z_0 = z_0^\rho = (323 \ {\rm MeV})^{-1} $,} 
 experimental
$f_\pi$ is obtained for $ \alpha = (424\,{\rm MeV})^3 $. For these values,  the argument $a   \equiv \alpha z_0^3$
of the modified Bessel functions in Eq.~(\ref{fpi2})  equals \mbox{$2.26\equiv a_0$.} Since $I_{2/3} (a) /
I_{-2/3} (a) \approx 1$ for $a \gtrsim 1$, then 
\begin{align}
\label{fpias}
{g_5^2}  f_\pi^2 \approx  3\cdot 2^{1/3} \, \frac{\Gamma (2/3) }{\Gamma (1/3) }\,
  {\alpha^{2/3}}  \ , 
\end{align}
i.e., the value of $f_\pi$ is  basically   determined by $\alpha$ alone (the same observation was made in the pioneering paper \cite{DaRold:2005zs}
and in a recent paper \cite{Kwee:2007dd}  in which the pion channel was studied numerically).
\begin{widetext}
In Fig.~\ref{psiphi}, we illustrate the behavior of the pion wave functions $\Psi$ and $\Phi$ representing them
 $\Psi \to  
\psi (\zeta,a)$, $\Phi \to \varphi (\zeta,a)$
as functions of dimensionless variables $\zeta \equiv z/z_0$  and $a \equiv \alpha z_0^3$.
For $a=0$ (i.e. when the chiral symmetry breaking parameter $\alpha$ vanishes), 
the limiting forms are  $\psi (\zeta ,0)= 1$ and
 $\varphi(\zeta ,0)= 1-\zeta^4$ \cite{Grigoryan:2007wn}.  As $a$ increases, both functions
become  more and more narrow, with $\psi (\zeta,a)$ becoming smaller and smaller at 
the IR boundary $\zeta =1$.

\begin{figure}[h]
\includegraphics[width=3in]{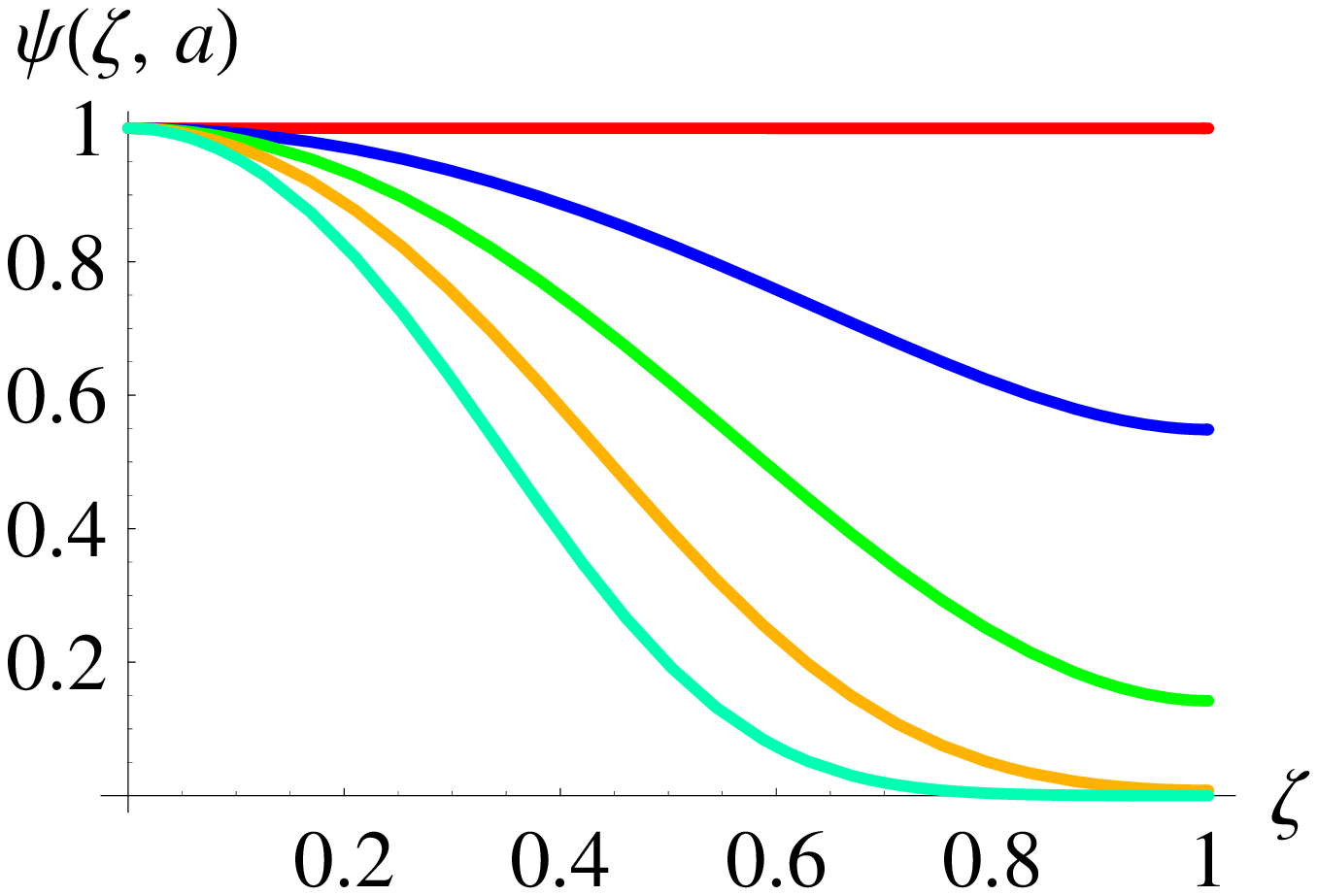} \includegraphics[width=3in]{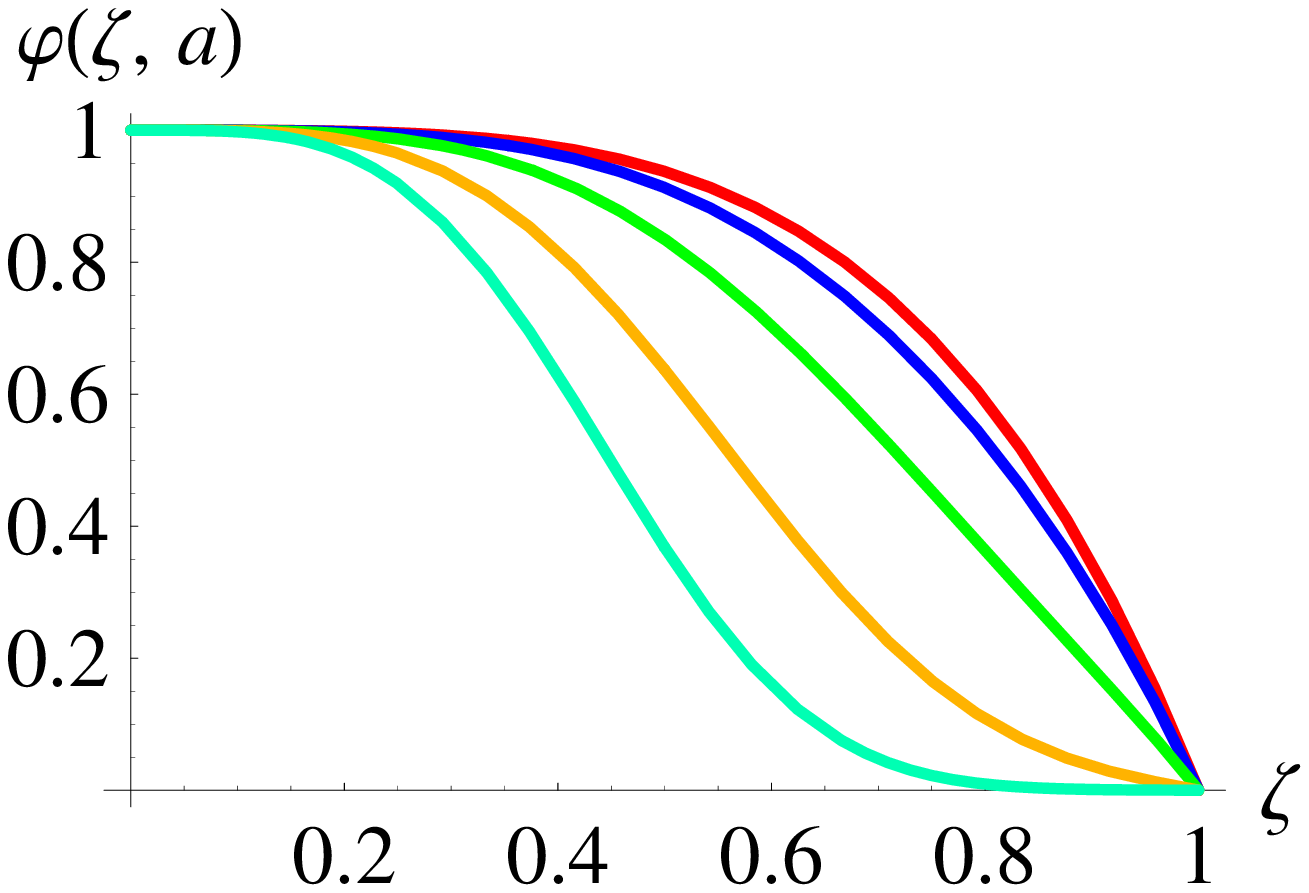} 
\caption{\label{psiphi}
Functions   $\psi(\zeta ,a)$   (left) and $\varphi(\zeta ,a)$ (right)
 for several values of $a$:
$a=0$ (uppermost lines), $a=1$, $a=2.26$, $a=5$, $a=10$ (lowermost lines).
}
\end{figure}
 \end{widetext}

\section{Anomalous amplitude}

\subsection{Isosinglet fields}

The $\pi^0 \gamma^* \gamma^*$ form factor is defined  by
\begin{eqnarray}
\label{ff}
\int
\langle {\pi}, {p}
|T\left\{J^{\mu }_{\rm EM}(x)\,J^{\nu}_{\rm EM}(0)\right\}| 0 \rangle e^{-iq_1 x } d^4 x \nonumber \\
 = \epsilon^{\mu  \nu \alpha  \beta}
q_{1 \, \alpha} q_{2\, \beta} \, F_{\gamma^*\gamma^*\pi^0} \left(Q_1^2,Q_2^2 \right ) ,\label{eq:form}
\end{eqnarray}
where $ p = q_1 + q_2 $ and $ q^2_{1,2} = -Q^2_{1,2} $. Its  value  for real photons
\begin{align}
F_{\gamma^* \gamma^* \pi^0}(0,0) = \frac{N_c}{12 \pi^2 f_\pi}
\end{align}
is  related in QCD  to the axial anomaly.

Since  $\pi^0$  meson is described by the third component $A^3$ of an  isovector field,  only the isovector
component of the product of two electromagnetic currents gives nonzero contribution  to the matrix element in
Eq.~(\ref{ff}). In   QCD,  the electromagnetic current is   given   by the sum
\begin{align}
J^{{\rm EM}}_{\mu} = J_{\mu}^{\{I=1\}, 3}+ \frac{1}{3} \, J_{\mu}^{\{I=0\}}
\end{align}
of isovector $J_{\mu}^{\{I=1\}, a}(x) $  (``$\rho$-type'') and isosinglet  $ J_{\mu}^{\{I=0\}}(x) $
(``$\omega$-type'')   currents,
\begin{align}
J_{\mu}^{\{I=1\}, 3}&= \frac{1}{2}\left(\bar{u} \gamma_{\mu} u
- \bar{d} \gamma_{\mu} d \right) = \bar{q}
\gamma_{\mu}t^3 q  \ ,
\\ \nonumber
J_{\mu}^{\{I=0\}} &= \frac{1}{2}\left (\bar{u} \gamma_{\mu} u + \bar{d} \gamma_{\mu} d \right)= \frac{1}{2}\bar{q}
\gamma_{\mu} \mathbb{1}\, q \ .
\end{align}
As a result,  the  matrix element
$\langle \pi^0|  J^{{\rm EM}} J^{{\rm EM}} | 0 \rangle$
is  nonzero  since it contains
$\langle \pi^0|   J^{\{I=1\}, 3} J^{\{I=0\}}  |0 \rangle \sim {\rm Tr} (t^3 t^3)$ parts.

To  extract   $F_{\gamma^*\gamma^*\pi^0}(Q_1^2,Q_2^2) $   form  factor
{\it  via}   holographic  correspondence,
we consider  the correlator of  the axial-vector current
 \begin{align}
J_{\mu}^{A,a}= \bar{q} \gamma_{\mu}\gamma_5 t^a q
\end{align}
and  vector  currents $J_{\mu}^{\{I=1\}, b}$, $ J_{\nu}^{\{I=0 \}}$. To proceed, we need to have isoscalar fields
on the AdS side of the holographic correspondence. This is achieved by gauging $U(2)_L \otimes U(2)_R $ rather
than $SU(2)_L  \otimes SU (2)_R$ group in the AdS/QCD action, i.e., by substituting  $t^a B^a_\mu$ by
\begin{align}
{\cal B}_{\mu} = t^a   B^a_\mu +   \mathbb{1} \, \frac{{\hat B}_\mu }{2} 
\end{align}
in Eq.~(\ref{AdS}). Then the 4D currents correspond to  the following 5D gauge fields
\begin{align}
J_{\mu}^{A,a}(x) & \rightarrow  A_{\mu}^a(x,z) \ ,  \\ \nonumber
J_{\mu}^{\{I=0\}}(x)  & \rightarrow {\hat V}_{\mu}(x,z) \ , \\ \nonumber
J_{\mu}^{\{I=1\}, a}(x) & \rightarrow  V_{\mu}^a(x,z) \ .
\end{align}

\subsection{Chern-Simons term}

The terms contained in the original AdS/QCD action (\ref{AdS}) cannot produce a 3-point function accompanied by 
the Levi-Civita $ \epsilon^{\mu  \nu \alpha  \beta}$  factor. 
However, such a contribution may be obtained by  adding the Chern-Simons
(CS) term. We follow Refs.~\cite{Witten:1998qj, Hill:2006wu} in choosing the  form of the $ {\cal O} (B^3) $ part
of the $ D = 5 $ CS action (the only one we need).  In the axial gauge $B_z = 0$, it is  written   as
\begin{align}
\label{cs}
S^{(3)}_{\rm CS}[{\cal B}] = k\, \frac{N_c}{48\pi^2}\epsilon^{\mu\nu\rho\sigma}&
{\rm Tr} \int d^4 x\, dz  \left(\partial_z
{\cal B}_{\mu}\right)
\nonumber \\ & \times
\biggl[{\cal F}_{\nu \rho}{\cal B}_{\sigma} + {\cal B}_{\nu}{\cal F}_{\rho \sigma} \biggr] ,
\end{align}
where ${\cal F} $  is the field-strength generated by ${\cal B}$
and $k$ should be  an integer  number adjusted
to  reproduce  the QCD anomaly result.
Then, in the  $U(2)_L \otimes U(2)_R $ model,  the relevant part of the CS term  is
\cite{Domokos:2007kt}
\begin{align}
S^{\rm AdS}_{\rm CS}[{\cal B}_L, {\cal B}_R] = S^{(3)}_{\rm CS}[{\cal B}_L]  - S^{(3)}_{\rm CS}[{\cal B}_R]  \ .
\end{align}
Taking into account that $ {\cal B}_{L,R} = {\cal V} \pm {\cal A} $, and keeping only
the  longitudinal component  of  the axial-vector field $ A \to  A_{\parallel} $
(that brings in the pion),  for which $ F^A_{\mu \nu} = 0 $,
we  have
\begin{eqnarray} \label{CSor}
S^{\rm AdS}_{\rm CS} &=&k \, \frac{N_c}{24\pi^2}
\epsilon^{\mu\nu\rho\sigma}\int d^4 x \int_0^{z_0}  dz \\ \nonumber
 &\times& \biggl[ \left(\partial_{\rho}{V}_{\mu}^a\right)
  \left( A^a_{\parallel \sigma}\stackrel{\leftrightarrow}{\partial_z } {\hat V}_{\nu}\right)
 +  \left(\partial_{\rho} {\hat V}_{\mu}\right)
 \left( A^a_{\parallel \sigma}\stackrel{\leftrightarrow}{\partial_z }V^a_{\nu} \right)
\biggr]  \, ,
\end{eqnarray}
where $\stackrel{\leftrightarrow}{\partial_z }\,  \equiv\,  \stackrel{\rightarrow}{\partial_z } - \stackrel{\leftarrow}{\partial_z }$.
Representing  $ A^a_{\parallel \sigma} =\partial_\sigma \psi^a$ and integrating by parts
gives
\begin{eqnarray}
\label{actionPhi}
S^{\rm AdS}_{\rm CS} &=&k\, \frac{N_c}{24\pi^2}\epsilon^{\mu\nu\rho\sigma}
\int d^4 x \int_0^{z_0}  dz \\ \nonumber
 &\times& \biggl[ 2 (\partial_z  \psi^a) \left( \partial_\rho  V^a_{\mu} \right)   \left(\partial_{\sigma} {\hat V}_{\nu}\right)
- \psi^a \partial_z  \left(\partial_\rho  V^a_{\mu} \, \partial_{\sigma} {\hat V}_{\nu}
  \right)
\biggr]  \, .
\end{eqnarray}
Note that  the $z$-derivative of $\psi^a$ is  proportional
to the $\Phi$-function of the pion (\ref{Phi}), which, as we argued in Refs.~\cite{Grigoryan:2007vg,Grigoryan:2007my,Grigoryan:2007wn},
is  the most direct analog of the quantum-mechanical
bound state wave functions, i.e., it  is the derivative $\partial_z\psi^a(z)$
rather than $\psi^a(z)$  itself that  is an   analog of the pion 4D  field. 
Then, the first term in the square brackets
in Eq.~(\ref{actionPhi}) has the structure similar  to the $ \pi \omega \rho $
interaction term
\begin{align}\label{HLS}
{\cal L}_{\pi \omega \rho} = \frac{N_c g^2}{8\pi^2 f_{\pi}}\epsilon^{\mu\nu\alpha\beta}{ \pi}^a \,
(\partial_{\mu}{ \rho}_{\nu}^a)\, \left(\partial_{\alpha}{ \omega}_{\beta}\right) \ ,
\end{align}
obtained in the hidden local symmetries approach \cite{Fujiwara:1984mp} from the anomalous
Wess-Zumino \cite{Wess:1971yu,Witten:1983tw}  lagrangian 
(see also a review \cite{Meissner:1987ge}). Here $ g $ is the universal gauge coupling
constant of that   approach. 

Integrating by parts over $z$,  the second term of Eq.~(\ref{actionPhi}) can be also converted into a contribution
of this form plus  a  $z=z_0$   surface term: 
\begin{eqnarray}
\label{actionPhi2}
S^{\rm AdS}_{\rm CS} &=&k\, \frac{N_c}{24\pi^2}\epsilon^{\mu\nu\rho\sigma}
\int d^4 x  \biggl  \{ -  \biggl [ \psi^a   (\partial_\rho  V^a_{\mu}) \, (\partial_{\sigma} {\hat V}_{\nu} ) \biggr ]
\biggr |_{z=z_0}   \nonumber  \\
 &+&3 \int_0^{z_0}   dz \,  (\partial_z  \psi^a) \left( \partial_\rho  V^a_{\mu} \right)   \left(\partial_{\sigma} {\hat V}_{\nu}\right)
\biggr \}   \, .
\end{eqnarray}
  The  latter can be eliminated by adding  a compensating surface term into the
original Chern-Simons  action, so that the resulting anomalous part of the action in the extended AdS/QCD model
\begin{eqnarray}
\label{actionPhi3}
S^{\rm  anom} & =&k\, \frac{N_c}{8\pi^2}\epsilon^{\mu\nu\rho\sigma}
\int d^4 x 
 \nonumber  \\ &\times&  
  \int_0^{z_0}   dz \,  (\partial_z  \psi^a) \left( \partial_\rho  V^a_{\mu} \right)   \left(\partial_{\sigma}\hat{V}_{\nu}\right)
\end{eqnarray}
has the structure of  Eq.~(\ref{HLS}).


\subsection{Three-point function}

The action (\ref{actionPhi2}) produces the  3-point function
\begin{align}
&\langle J_{\alpha}^{A,3}(-p)J^{\rm EM}_{\mu} (q_1)J^{\rm EM}_{ \nu}(q_2) \rangle \\
\nonumber &= T_{\alpha
\mu \nu}(p,q_1,q_2) \ i(2\pi)^4 \delta^{(4)}(q_1 + q_2 - p) \ ,
\end{align}
with
\begin{align}\nonumber
T_{\alpha \mu \nu}(p,q_1,q_2) = & \frac{N_c}{12 \pi^2}\frac{p_{\alpha}}{p^2} \,
\epsilon_{ \mu \nu  \rho \sigma} \, q_{1}^\rho q_{2}^{\sigma}K_b(Q^2_1,Q^2_2) \ .
\end{align}
Here, $ p $ is the momentum of the pion and $ q_{1}, q_{2} $ are the momenta
of the  photons.

The ``bare''  function $K_b (Q_1^2,Q_2^2)$  is   given  by
\begin{eqnarray}
\label{k0qq}
K_b(Q_1^2,Q_2^2)  =- \frac{k}{2}   \int_0^{z_0}  {\cal J}(Q_1,z) {\cal J}(Q_2,z) \partial_z \psi (z) \, dz \, ,
\end{eqnarray}
where  ${\cal J} (Q_{1},z)$,  ${\cal J} (Q_{2},z)$  are the vector bulk-to-boundary propagators
(\ref{JQz}),  and $\psi (z)$ is the pion wave function (\ref{Psi}).

\subsection{Conforming to QCD axial  anomaly}

For real photons, i.e., when $Q_1^2=Q_2^2=0$,  the  value  of the
$F_{\pi \gamma^* \gamma^*}(Q_1^2,Q_2^2) $
form factor   in QCD 
(with massless quarks) is settled by the axial 
anomaly, which   corresponds to  $K^{\rm QCD} (0,0)=1$.
Our goal is to build an  AdS/QCD model  for $F_{\pi \gamma^* \gamma^*}(Q_1^2,Q_2^2) $
that  reproduces  this 
value.
Taking  $Q_1^2=Q_2^2=0$, we have ${\cal J}(0 \, ,z) =1$,  and Eq.~(\ref{k0qq}) gives
\begin{align}
\label{1-Psiz0}
K_b(0,0) = -  \frac{k}{2}  \int_0^{z_0} \partial_z \psi (z) \, dz  = &  -   \frac{k}{2} \,  \psi (z_0)
\nonumber \\
= & \frac{k}{2} \, \biggl  [1  - \Psi (z_0)  \biggr]  \  .
\end{align}
On the IR boundary $z=z_0$, the pion wave function $\Psi (z)$ from  Eq.~(\ref{Psi})
is 
\begin{align}
\Psi (z_0 ) =
\frac{\sqrt{3} \,  \Gamma \left ({2}/{3} \right  )}{\pi I_{-2/3}(a)}
\left(\frac{1}{2a^2}\right)^{1/3}  \ .
\end{align}
 As we discussed  above, experimental values of $m_\rho$ and $f_\pi$ correspond to
$a=2.26$, which gives $\Psi (z_0 )=0.14$. 
The  magnitude of  $\Psi (z_0)$ rapidly decreases for larger $a$ (e.g.,
$\Psi (z_0)|_{a=4} \approx 0.02$, see Fig.~\ref{psiphi}). 
Still,  
the value of  $\Psi (z_0)$ is   nonzero 
at any finite value of $a$, and it is impossible 
to exactly reproduce the anomaly  result by simply 
adjusting the integer number $k$. 
To conform to the QCD anomaly value  $K^{\rm   QCD}(0,0)=1$, 
we  add  a  surface term compensating 
the $\Psi(z_0)$ contribution in Eq.~(34) and then take $k=2$.
To fix the  form  of the surface term, we note first that it should have the structure 
of a  $VVA$ 3-point function taken on the 
$z=z_0$  surface.  Furthermore, using the derivatives ${\cal J}'_z  (Q_i,z_0)$
of the bulk-to-boundary propagators in this term  is excluded,
because ${\cal J}'_z  (Q_i=0,z_0) =0$ at the real photon point.   
On the other hand,  
${\cal J}  (Q_i=0,z_0) =1$,  and 
 our final   model  for $F_{\pi \gamma^* \gamma^*}(Q_1^2,Q_2^2) $
corresponds to the function
\begin{eqnarray}
\label{kqqmod}
K (Q_1^2,Q_2^2)  &=&
\Psi (z_0)  {\cal J}(Q_1,z_0) {\cal J}(Q_2,z_0) 
	 \\ &-&    \int_0^{z_0}  {\cal J}(Q_1,z) {\cal J}(Q_2,z)\,   \partial_z \Psi (z) \, dz \ . \nonumber
\end{eqnarray}
The  extra   term   provides $K (0,0)=1$, 
and since 
\begin{align}
 {\cal J} (Q,z_0)= \frac1{I_0 (Qz_0)} \  , 
\end{align}
it   rapidly decreases with the growth of   $Q_1$ and/or  $Q_2$.
We will see later that in these regions the behavior 
of $F_{\pi \gamma^* \gamma^*}(Q_1^2,Q_2^2) $  is determined 
by small-$z$ region of integration. 
Thus, the effects of fixing the $\Psi (z_0) \neq 0$ 
artifact at  the infrared  boundary  
are wiped out in  the  ``short-distance''  regime.


\section{Momentum dependence}

\subsection{Small virtualities }

If one of the photons is  real $Q_1^2=0$, while  another is   almost real, $Q_2^2 = Q^2 \ll 1/z_0^2$, we may use
the expansion 
\begin{align}
{\cal J}(Q,z) =  1 - \frac{1}{4}\, Q^2z^2\left[1 -  \ln \frac{z^2}{z_0^2} \right] + {\cal O} (Q^4)  \ ,
\end{align}
which gives (for  $z_0=z_0^\rho$   and $a=a_0=2.26$)
\begin{align}
\label{kslope}
{K(0, Q^2)}\simeq & \, 1  - 0.66 \, \frac{Q^2 z^2_0}{4}
\nonumber \\  \simeq & \, 1  - 0.96 \, \frac{Q^2}{m_\rho^2} \ .
\end{align}
The predicted slope is very close to  the value $1/m_\rho^2$ 
expected from a  naive vector-meson  dominance. Experimentally, the slope of the  $F_{\pi \gamma^*
\gamma^*}(0,Q^2) $ form factor for small timelike (negative) $Q^2$ is  measured through the Dalitz decay \mbox{$\pi^0
\to e^+e^- \gamma$.}  In our notations, the  usual  representation of the results is 
\begin{align}
K(0,Q^2) = 1- a_\pi \, \frac{Q^2}{m_{\pi}^2}  \  ,
\end{align}
where $m_{\pi}$  is the experimental pion mass.
Then the   $Q^2$-slope given   by Eq.~(\ref{kslope}) corresponds to \mbox{$a_\pi \approx 0.031$.} This number is
not very far from the central values  of two last experiments,  $a_\pi =0.026 \pm 0.024 \pm 0.0048$
\cite{Farzanpay:1992pz},
 $a_\pi =0.025 \pm 0.014 \pm 0.026$ \cite{MeijerDrees:1992qb},
but the experimental errors are rather  large. 
An earlier  experiment \cite{Fonvieille:1989kj}  
produced   $a_\pi = -0.11  \pm 0.03 \pm 0.08$, a  result whose central value
has opposite sign and much larger absolute magnitude.
 In  the spacelike   region,  the data are available only 
for  the values   \mbox{$Q^2\gtrsim 0.5$\, GeV$^2$}  (CELLO~\cite{Behrend:1990sr})
and  \mbox{$Q^2\gtrsim 1.5$\, GeV$^2$}   (CLEO~\cite{Gronberg:1997fj})
which cannot  be treated as very small. 
The  CELLO collaboration~\cite{Behrend:1990sr}   gives   the value   \mbox{$a_\pi = 0.0326\pm 0.0026$}
that   is   very   close to our result. 
To settle the uncertainty of the timelike data  (and also on its own grounds),
it would be interesting to have data
on the slope from the spacelike   region of very   small $Q^2$,
which may be obtained by modification of the PRIMEX
experiment \cite{primex} at JLab.

\subsection{Large virtualities}

Since ${\cal J} (Q,z_0)$  exponentially $\sim e^{-Qz_0}$ vanishes
for large $Q$,   we   can   neglect the first term of Eq.~(\ref{kqqmod}) in the
asymptotic $Q^2 \to \infty$  region.
 Representing $\partial_z \psi (z)$ in   terms  of the
$\Phi (z)$   wave   function,
we   write $K(Q_1^2,Q_2^2) $ as
\begin{eqnarray}
\label{kQQ}
K(Q_1^2,Q_2^2) \simeq 
  \frac{s_0}{2}   \int_0^{z_0}  {\cal J}(Q_1,z)  {\cal J}(Q_2,z) \,  \Phi (z) \, z\, dz\ .
\end{eqnarray}
Our goal  is  to compare
the predictions based on this   formula
with the {\it  leading-order } perturbative QCD results.
Note that the situation is different from that with the
charged pion form factor, where the leading-power  $1/Q^2$   
pQCD result  \mbox{$F_\pi^{\rm pQCD} (Q^2) \to 2\, (\alpha_s/\pi) \,s_0/Q^2$ 
\cite{Efremov:1979qk}}  corresponding to  {\it hard}  contribution  is the ${\cal O} (\alpha_s) $ correction 
to the  {\it soft} contribution,  for which  AdS/QCD    gives  \mbox{$F_\pi^{\rm AdS/QCD} (Q^2) \to s_0/Q^2$
\cite{Grigoryan:2007wn}}. In that situation, 
 it makes no sense to discuss whether  pQCD and AdS/QCD asymptotic predictions
agree with each other  numerically  or not. 
In general,  our AdS/QCD  model contains no information about  hard gluon exchanges
of pQCD that produce the 
$\alpha_s$  factors. 
However, the 
pQCD expression for the 
\mbox{$\gamma^* \gamma^* \to \pi^0$}  form factor 
has {\it zero} order in $\alpha_s$,
so  now it makes sense to compare the 
{\it leading-order} pQCD predictions for this particular
(in fact, exceptional) form factor   with AdS/QCD calculations.

It is  instructive to consider first
two simple kinematic situations ($Q_1^2=0,  \, Q_2^2=Q^2$  and $Q_1^2=Q_2^2=Q^2$),
and then analyze the general case.

\subsubsection{One real photon}

Form factor in the kinematics when the virtuality of one of the photons can be treated as zero $Q_1^2 \approx 0$,
while  another $Q_2^2 =Q^2$  is large was studied experimentally by CELLO~\cite{Behrend:1990sr} and
CLEO~\cite{Gronberg:1997fj} collaborations.

In perturbative QCD, the   $K(0,Q^2)$  form factor  at large $Q^2$ is   obtained
 from the   factorization formula
\begin{equation}
\label{TpQCD}
 K(0,Q^2) = \int_0^1 T(Q^2,x) \, \varphi_\pi (x) \, dx \  ,
\end{equation}
where $ T(Q^2,x)$ is  the amplitude  of the hard subprocess $\gamma \gamma^* \to \bar q q$. The   latter,   modulo
logarithms of $Q^2$, has the $1/Q^2$ behavior. In   the lowest order, when
\begin{equation}
\label{K0QpQCD}
K^{\rm pQCD} (0,Q^2) = \frac{s_0}{3Q^2} \int_0^1 \frac{\varphi_\pi (x)}{x} \, dx
\equiv   \frac{s_0}{3Q^2} \, I^{\varphi} \  ,
\end{equation}
the  purely $1/Q^2$   outcome  reflects the large-$Q^2$  behavior of the hard quark propagator connecting the
photon vertices. A  particular form of the pion distribution amplitude (DA)  
is  irrelevant to the power of the large-$Q^2$
behavior of $ F(0,Q^2) $, as  far as it provides a convergent $x$-integral in Eq.~(\ref{K0QpQCD}). The latter
requirement is fulfilled, e.g.,  if the pion distribution amplitude $\varphi_\pi (x)$  vanishes at the end-points ($x=0$ or
1) as a positive power of $x(1-x)$.
Whether it vanishes  at $x=0$ as $x$, $x^2$ or $\sqrt{x}$  does not matter --  this would   not affect the
$1/Q^2$   large-$Q^2$ behavior of the $\gamma \gamma^* \pi^0 $ form factor in the lowest pQCD order. 
However, the
shape of the pion  distribution amplitude $\varphi_\pi (x)$ determines  the value of the  coefficient
$I^{\varphi}$ that provides the normalization  of the ${\cal O} (1/Q^2)$   term. For the asymptotic shape
$\varphi_\pi^{\rm as}  (x) = 6\, x (1-x)$, we have $I^{\varphi^{\rm as} } =3$  and $K^{\rm pQCD\, (as)} (0,Q^2) =
s_0/Q^2$.

Take now our extended AdS/QCD model for the $K (0,Q^2)$ form factor. It gives
\begin{eqnarray}
\label{k0Q}
K(0,Q^2) \simeq  
  \frac{s_0}{2}   \int_0^{z_0}  {\cal J}(Q,z)  \,  \Phi (z) \, z\, dz\ .
\end{eqnarray}
At first sight, this  expression, though completely different analytically from 
the   pQCD formula (\ref{TpQCD}),   has a  general structure similar to it: the
$Q^2$-dependence is  accumulated in the universal  current  factor  ${\cal J} (Q,z)$, while the  bound state
dynamics is  described by the \mbox{$Q^2$-independent}  wave function $\Phi (z)$. The  obvious difference is  that the
bulk-to-boundary propagator  ${\cal J} (Q,z)$   does not have a power behavior at   large $Q^2$: it  behaves  in
that region like $e^{-Qz}$, coinciding in this limit with  $ zQ K_1 (zQ)\equiv {\cal K} (Qz)  $, the free-field
version of the nonnormalizable   mode. The   {\it power behavior}  in $Q^2$ appears  only {after integration}  of
the {\it exponentially decreasing}  function  over $z$. As a  result, only  small values of $z$  are important in
the relevant integral, and  the outcome is determined by the small-$z$ behavior of the wave function $\Phi (z)$.
As   far as  $\Phi (z)$  tends to a  nonzero value $\Phi (0)$  when  $z \to 0$, the outcome is the $1/Q^2$
behavior:
\begin{equation}
\label{Fas}
 K  (0,Q^2)   \to \frac{  \Phi (0) s_0}{2 Q^2}
\int_0^\infty d\chi \, \chi^2 \,  K_1 (\chi) =  \frac{\Phi (0) \, s_0}{Q^2}  \ .
\end{equation}
Just like in the case of the (charged) pion EM form factor \cite{Grigoryan:2007wn},
the large-$Q^2$   behavior
of $ K(0,Q^2) $ is   determined by the   value of the $\Phi (z) $
wave function at  the origin.
Note, that this value   is   fixed:
$\Phi (0)= 1$, which
gives $ K(0,Q^2)  =  \, s_0/Q^2$,
the  result that  coincides with
the leading-order  prediction of pQCD
for  $I^{\varphi}=3$,
the value that is obtained, e.g., if one takes the
asymptotic pion DA $\varphi_\pi^{\rm as}  (x) = 6 x (1-x)$.

Experimentally,  the leading-order pQCD prediction
with $I^{\varphi}=3$  is  somewhat above the
  data.
However, the next-to-leading ${\cal O}(\alpha_s)$
pQCD correction is negative,
and decreases the result by about 15\%,
producing a satisfactory
agreement.
More elaborate fits  \cite{Bakulev:2007jv} favor
DAs  that differ from  the asymptotic one
by  higher Gegenbauer harmonics
$x(1-x) C_n (2x-1)$ with $n=2$ and $n=4$.
Still,  for all the  DAs obtained from these
fits,   the magnitude  of the integral
$
I^{\varphi } $  is  very close to  the value $I^{\rm as}=3$
given by the asymptotic DA  (see Ref.~\cite{Bakulev:2007jv} for details and
references).
Thus,  the result  of
 our  calculation
is  in full agreement with   the magnitude of the
leading-order pQCD  part of the  NLO fits of
existing experimental data.

\subsubsection{Equal virtualities}

Another interesting kinematics is when the photons  have equal large  virtualities,
 $Q_1^2=Q_2^2=Q^2$.
In this case,   the  leading-order pQCD prediction
\begin{equation}
\label{k12pQCD}
 K^{\rm pQCD} (Q^2,Q^2) = \frac{s_0}{3} \int_0^1 \frac{\varphi_\pi (x) \, dx}{xQ^2 + (1-x)Q^2} \,
= \frac{s_0}{3Q^2}
\end{equation}
does not depend on the shape of the pion DA. In our extended AdS/QCD model, we obtain
\begin{eqnarray}
K(Q^2,Q^2) \simeq 
  \frac{s_0}{2}   \int_0^{z_0}  [{\cal J}(Q,z)]^2   \Phi (z) \, z\, dz\ .
\end{eqnarray}
Asymptotically,  we have
\begin{equation}
\label{Fas2}
 K(Q^2,Q^2)   \to \frac{  \Phi (0) s_0}{Q^2}
\int_0^\infty d\chi \, \chi^3 \,  [K_1 (\chi)]^2 =  \frac{  s_0}{3Q^2}  \ ,
\end{equation}
which is the same result as in the leading-order   pQCD.

\begin{widetext}

\subsubsection{General case}

Finally,   let us   consider the general kinematics, when
\mbox{$Q_1^2 = (1+\omega)Q^2$,} and  $Q_2^2 = (1-\omega)Q^2$.
The leading-order pQCD formula gives in this case
\begin{align}
 K^{\rm pQCD} ((1+\omega)Q^2,(1-\omega)Q^2) = 
\frac{s_0}{3Q^2} \int_0^1 \frac{\varphi_\pi (x) \, dx}{1+\omega (2x-1)} 
\equiv
\frac{s_0}{3Q^2} \, I^\varphi (\omega) \  ,
\end{align}
while Eq.~(\ref{kQQ})  of our  AdS/QCD model reduces, for large $Q^2$,  to
 \begin{eqnarray}
\label{kQ1Q2}
K((1+\omega)Q^2,(1-\omega)Q^2)  \to  \frac{  \Phi (0) s_0}{2Q^2} \, \sqrt{1-\omega^2}
\int_0^\infty d\chi \, \chi^3 \,  K_1 ( \chi \sqrt{1+\omega})
 K_1 ( \chi \sqrt{1-\omega})  \equiv  \frac{  s_0}{3Q^2} \,  I^{\rm AdS}  (\omega) \,   \ ,
\end{eqnarray}
\end{widetext}
with  the function $I^{\rm AdS}  (\omega) $   given by
 \begin{eqnarray}
I^{\rm AdS}  (\omega) = \frac3{4\omega^3} \, \left [ 2 \omega - (1-\omega^2) \,
\ln \left ( \frac{1- \omega}{1+\omega} \right ) \right ] \  .
\end{eqnarray}
It is   straightforward  to check that $I^{\rm AdS}  (\omega) $ coincides with the pQCD function
$I^\varphi (\omega)$ calculated for the asymptotic
distribution amplitude $\varphi^{\rm as}  (x) = 6x(1-x)$.
Indeed,  using the representation
\begin{equation}
 \chi  K_1 ( \chi ) = \int_0^\infty e^{-\chi^2/4u -u} \, du \  ,
\end{equation}
we can easily integrate over $\chi$ in Eq.~(\ref{kQ1Q2}) to get
\begin{eqnarray}
\label{KQ1Q2}
K(Q_1^2, Q_2^2)  \to  \frac{  s_0}{Q^2} \,
\int_0^\infty  \int_0^\infty \frac{u_1 u_2 \,e^{-u_1-u_2}  du_1 du_2}{u_2 (1+\omega)  +u_1(1-\omega)}
  \,  .
\end{eqnarray}
Changing variables $u_2=x\lambda$, $u_1= (1-x) \lambda$
and integrating over $\lambda$, we obtain
\begin{align}
 K(Q_1^2, Q_2^2)  \to  &
\frac{s_0}{3Q^2} \int_0^1 \frac{6 \, x (1-x) \, dx}{1+\omega (2x-1)} \ ,
\end{align}
which coincides with the pQCD formula (\ref{k12pQCD})
if we take $\varphi_\pi (x) = 6\, x(1-x)$.

Note that the absolute normalization of the 
 asymptotic behavior of $K(Q_1^2, Q_2^2)  $ for large $Q_i$
in our model is  fixed by the choice $k=2$ that allows to
conform to the value $K(Q_1^2, Q_2^2) =1$ corresponding to the
QCD anomaly result.  As we have seen, this choice exactly reproduces  also 
the leading-order pQCD result for the equal-virtualities form factor $K(Q^2, Q^2)  $. 
The origin of this  rather unexpected result  needs further studies. 
Recall  also that in  pQCD the result for $K(Q^2, Q^2)  $ 
is the same for any pion distribution amplitude,
while the result  for  the 
unequal-virtualities form factor $K((1+\omega)Q^2,(1-\omega)Q^2)$
depends  on the  choice  of the pion distribution amplitude.
The fact that our AdS/QCD model  gives the same result 
as the leading-order pQCD 
calculation performed for the asymptotic distribution amplitude,
also  deserves a further  investigation.


\subsubsection{From small to large $Q^2$}

Both $K(0,Q^2)$  and $K(Q^2,Q^2)$   functions  
are \mbox{equal to 1 } at $Q^2=0$.  For  large $Q^2$,  the first one 
tends to $s_0/Q^2$ and the second one  to $s_0/3Q^2$. 
The  question is   how these functions interpolate between 
the  regions of small  and large $Q^2$.
Long ago, Brodsky and Lepage \cite{Brodsky:1981rp} proposed  
a simple monopole (BL) interpolation 
\begin{align}
\label{extrap}
 K^{\rm BL} (0,Q^2) = \frac1{1+Q^2/s_0}
\end{align}
between the $Q^2=0$ value  and the large-$Q^2$ asymptotic prediction of perturbative QCD
for $K(0,Q^2)$.
 Later, this  behavior was
obtained within the ``local quark-hadron duality''  approach \cite{Nesterenko:1982gc,Radyushkin:1995pj},  
in which $K(0,Q^2)$ is obtained by integrating
the spectral density $\rho (s, 0,Q^2)=Q^2/(s+Q^2)^2$ of the 3-point function over the ``pion duality interval''
$0\leqslant s \leqslant s_0$. The  curve for $Q^2 K(0,Q^2)$ based on Eq.~(\ref{k0Q}) 
practically coincides with  that
based  on  BL  interpolation (\ref{extrap}), see Figs.~\ref{K0Q} and \ref{K0Q2}.  
For    comparison, we also show on Fig.~{\ref{K0Q}    the monopole fit $K^{\rm CLEO} (0,Q^2) =1 /(1+Q^2/\Lambda_\pi^2)$ 
(with $\Lambda_\pi= 776$ MeV) 
of CLEO data  \cite{Gronberg:1997fj}.  As we mentioned, an accurate fit to CLEO data 
 \cite{Bakulev:2007jv} 
was  obtained in  the next-to-leading  order (NLO) pQCD, with the 
leading order part of the NLO pQCD fit being very close to BL-interpolation curve,
and hence, to our AdS/QCD result as well.
\begin{figure}[h]
\mbox{
\includegraphics[width=3in]{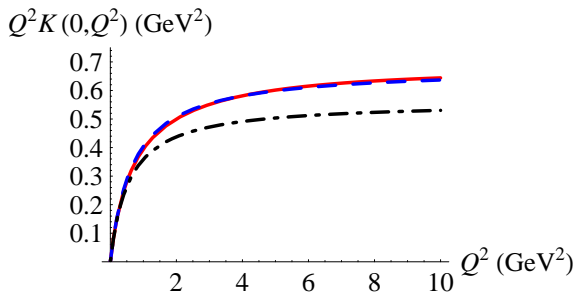} \hspace{1cm}}
\caption{\label{K0Q}
Function  $Q^2 K(0,Q^2)$  in AdS/QCD model (solid curve,  red online) and 
in local quark hadron  duality model (coinciding with Brodsky-Lepage
interpolation formula, dashed curve, blue  online). 
The monopole fit of  CLEO data is   shown by dash-dotted curve (black  online).
}
\end{figure}

\begin{figure}[h]
\includegraphics[width=3in]{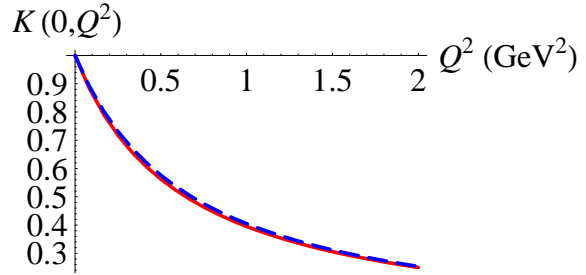}  \hspace{1cm}
\caption{\label{K0Q2}
Form factor $K(0,Q^2)$ in AdS/QCD model (solid curve, red online) compared to 
BL interpolation  formula (dashed curve, blue online).
}
\end{figure}

In  the  case of $K(Q^2,Q^2)$ function,   our model predicts the slope $1.92/m_\rho^2 \approx 2.15/s_0$
(twice  the slope of $K(0,Q^2)$, see Eq.~(\ref{kslope})),
while the local duality model  gives   \cite{Radyushkin:1995pj}
\begin{align}
& K^{\rm LD} (Q^2,Q^2)= 1 - Q^4 \int_0^1 \frac{dx}{[Q^2 +s_0 x (1-x)]^2} \nonumber \\
=& 1- \frac{2Q^2}{s_0+4 Q^2} -\frac{8Q^4 \tanh^{-1} \sqrt{s_0/(s_0+4Q^2)}}
{\sqrt{s_0 (s_0+4Q^2)^3} }
\  ,
\end{align}
a   curve which has the slope  $2/s_0$ at $Q^2=0$ :  
\begin{align}
 K^{\rm LD} (Q^2,Q^2) =1- 2\,  \frac{Q^2}{s_0} + {\cal O} (Q^4) \   .
\end{align}
However,   higher terms of $Q^2$ expansion 
become important for $Q^2$ as small as 0.01 GeV$^2$,
where  $K^{\rm AdS} (Q^2,Q^2)$ becomes  larger than 
$K^{\rm LD} (Q^2,Q^2)$, and the ratio $K^{\rm AdS} (Q^2,Q^2)/K^{\rm LD} (Q^2,Q^2)$
reaches its maximum value of 1.08 for $Q^2 \sim 0.3\,{\rm GeV}^2$,
then slowly decreasing towards the limiting value of 1. 
For large $Q^2$,  the  local duality model  gives the same result 
\begin{align}
 K^{\rm LD} (Q^2,Q^2) =\frac{s_0}{3Q^2} + {\cal O} (1/Q^4) 
\end{align}
as our  present model  (\ref{Fas2})  and pQCD   (\ref{k12pQCD}).
As a consequence, our  present model   produces a curve 
that is   very close to     the  curve based on the local duality model,
see Fig.~\ref{figK0Q}.
\begin{figure}[h]
\includegraphics[width=3in]{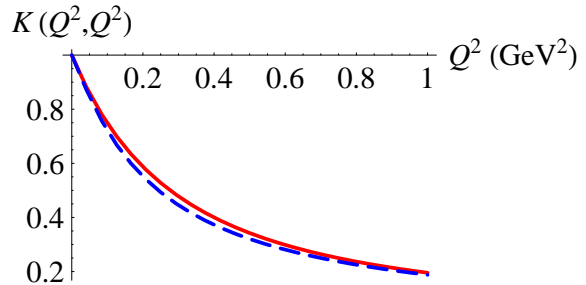}   \hspace{1cm}
\caption{\label{figK0Q}
Form factor $K(Q^2,Q^2)$ in AdS/QCD model (solid curve, red online) compared to the 
local   quark-hadron   duality model prediction  (dashed curve, blue online).
}
\end{figure}

\section{Bound-state  decomposition}

The bulk-to-boundary propagator ${\cal J} (Q,z)$
may be written as a sum
\begin{align}
\label{GVMD} 
 {\cal J} (Q,z)  = \sum_{n = 1}^{\infty}\frac{  g_5f_{n} \psi_n^V  ( z)}{ Q^2 + M^2_{n} }
\end{align}
over all vector   bound states.  For this reason,
the form factor $K(Q_1^2,Q_2^2)$
also has  a  generalized vector meson dominance (GVMD) representation
 \begin{align}
\label{Ank}
  K(Q_1^2,Q_2^2) = \sum_{n=1}^{\infty}  \sum_{k=1}^{\infty}  \frac{A_{n,k}+B_{n,k}}{(1+Q_1^2/M_n^2)
(1+Q_2^2/M_k^2)} \ ,
 \end{align}
where  $A_{n,k}$'s  come  from the first (surface)  term in 
Eq.(\ref{kqqmod}),
 \begin{align}
  A_{n,k} = \frac{4 \Psi (z_0)}{\gamma_{0,n} \gamma_{0,k}J_1(\gamma_{0,n})J_1(\gamma_{0,k})} \ ,
 \end{align}
while $B_{n,k}$'s are obtained from  the   second term and are given by
the  convolutions
\begin{align}
  B_{n,k}  &= -\frac{4}{z_0^2 \gamma_{0,n}\gamma_{0,k}  J_1^2 (\gamma_{0,n})J_1^2(\gamma_{0,k})} 
\nonumber \\ &  \times \int_0^{z_0}
  \Psi' (z) \, J_1 (\gamma_{0,n}z/z_0) \, J_1 (\gamma_{0,k}z/z_0) \, z^2 
\, dz  \  .
 \end{align}
Let us  study  the  structure of the bound-state decomposition in two
most interesting cases: for $K(0,Q^2)$ and $K(Q^2,Q^2)$.

\subsection{One real photon}

To study the bound-state decomposition  of $K(0,Q^2)$,
we write  the basic
expression
 \begin{align}
  K(0,Q^2) =  \Psi (z_0) {\cal J} (Q,z_0) - \int_0^{z_0}  {\cal J} (Q,z) \, \Psi'(z)  \, dz
 \end{align}
and use GVMD representation (\ref{GVMD})  for $  {\cal J} (Q,z)$. This  gives
 \begin{align}
\label{GVMDK0}
  K(0,Q^2) =   \sum_{n=1}^{\infty} \frac{A_n+B_n}{1+Q^2/M_n^2} \ ,
 \end{align}
where 
 \begin{align}
  A_n =  \frac{2 \Psi (z_0)}{\gamma_{0,n} J_1 (\gamma_{0,n})}
 \end{align}
and 
 $B_n$'s  coincide with    the coefficients
 \begin{align}
\label{Bn}
  B_n = -\frac{2}{z_0 \gamma_{0,n} J_1^2 (\gamma_{0,n})} \int_0^{z_0}
  \Psi' (z) \, J_1 (\gamma_{0,n}z/z_0) \,  z
\, dz
 \end{align}
for  the expansion
\begin{align}
  \Psi' (z) =  - \frac1{z_0} \sum_{n=1}^{\infty} B_n  \gamma_{0,n}  J_1 (\gamma_{0,n}z/z_0)
 \end{align}
of the pion wave function $\Psi' (z)$ over the $\psi_n^V (z)/z$  wave functions
(\ref{phi})
of vector meson bound states.
In particular,
 \begin{align}
  K(0,0) &= \sum_{n=1}^{\infty}{A_n } + \sum_{n=1}^{\infty}{B_n }\nonumber \\
&= \Psi (z_0) +[\Psi(0) - \Psi (z_0)] = 1  \  .
 \end{align}
This relation may  be directly obtained from the formula
\begin{align}
\left. 2 \sum_{n=1}^{\infty} \frac{x J_1(\gamma_{0,n} x)}
{\gamma_{0,n}J^2_1(\gamma_{0,n})} \right |_{x \leqslant 1}  = 1  \ .
\end{align}
%

The bound state   decomposition of $K(0,0)$ looks like
 \begin{align}
\label{slow}
K(0,0) = 1=0.9512+& 0.0408+0.0446-0.0753 \\ 
+& 0.0764-0.0736+0.0703+ \ldots \  . \nonumber 
 \end{align}
There is a  strong dominance of the lowest vector state, while each of the higher states is 
suppressed by more than factor of 10.
The slow convergence of higher terms is  due to $A_n$ terms proportional 
to $\Psi (z_0)\approx 0.14$. For large $n$, one can approximate
$A_n \approx  \Psi(z_0) (-1)^n \sqrt{2/n}$.

Integrating by parts in Eq.~(\ref{Bn})  gives a representation 
directly for the  total coefficient
 \begin{align}
  A_n +B_n = \frac{2}{z_0^2 J_1^2 (\gamma_{0,n})} \int_0^{z_0}
  \Psi (z) \, J_0 (\gamma_{0,n}z/z_0) \,  z
\, dz  \  , 
 \end{align}
that  is related to the expansion of the pion wave function $\Psi (z)$ over the $\phi_n$-functions
(\ref{phi}) 
of vector meson bound states:
 \begin{align}
  \Psi (z) =  \sum_{n=1}^{\infty} (A_n +B_n)\,  J_0 (\gamma_{0,n}z/z_0) \  .
 \end{align}
Using it, one obtains again that  $K(0,0) = \Psi (0) = 1$. 
The slow convergence of higher terms in Eq.~(\ref{slow})  is now  explained by the 
necessity to reproduce the 
finite value of $\Psi (z)$  at $z=z_0$
by functions vanishing at $z=z_0$.

The slope of $K(0,Q^2)$ at $Q^2=0$ is given
by the sum of $(A_n+B_n)/M_n^2$ coefficients, which converges   rather  fast:
 \begin{align}
K(0,Q^2)=1 -& \frac{Q^2}{m_\rho^2} \biggl \{0.9512 + 0.0077+0.0034-0.0031  \nonumber  \\ &
 +  0.0020
-0.0013  +0.0009 + \ldots \biggr \} \ ,
 \end{align}
and  the contribution of the lowest state completely
dominates the outcome.

Each term of the GVMD expansion (\ref{GVMDK0})
behaves like $1/Q^2$ at large $Q^2$.
In particular, the lowest-state contribution
behaves like $0.95\, m_\rho^2/Q^2 \approx (0.57\,{\rm GeV}^2)/Q^2$.
We also obtained above that $K(0,Q^2)$
behaves asymptotically like $s_0/Q^2 \approx (0.67\,{\rm GeV}^2)/Q^2$.
The two scales are not   so   different, and one may be tempted
to speculate  that the large-$Q^2$ behavior of $K(0,Q^2)$
also reflects the dominance of the lowest resonance.
However,  the coefficient of $1/Q^2$ is  formally
given by the sum of $(A_n+B_n)M_n^2$  terms, which does  not  show
good   convergence even after 7 terms are taken:
 \begin{align}
  \sum_{n=1}^\infty & (A_n+B_n) M_n^2 = 
  m_\rho^2 \biggl \{0.951+0.215+0.577  \nonumber \\ &
 -1.811 +
2.945-4.158+5.473 + \ldots \biggr \} \ .
 \end{align}


A  simple comment is in order now: within the AdS/QCD model \cite{Erlich:2005qh}
the
$\rho$-meson mass is determined by the ``confinement radius''
$z_0$, while
the scale  $s_0=8\pi^2 f_\pi^2$
is basically determined by the chiral symmetry breaking
parameter $\alpha$ (see Eq.~(\ref{fpi2}) and preceding discussion).
Calculationally, the coefficient $s_0$ of $1/Q^2$ asymptotic behavior
was determined solely by the magnitude 
 of the pion  wave function  $\Psi' (z) /z $ at the origin.
Furthermore,  it was legitimate to take  the free-field
form of the vector bulk-to-boundary propagator in our
calculation, i.e., no  information about vector channel
mass scales was involved.

Moreover, one may  write the bound-state decomposition
for the $J(Q,z_0)$ function.   Again, each term of such a decomposition 
has $1/Q^2$ asymptotic behavior, while 
$J(Q,z_0)$ {\it exponentially} decreases for large $Q$.
In fact,  the formal  sum $\sum_n A_n M_n^2$ in this  case
diverges like $\sum_n (-1)^n n^{3/2}$.

Summarizing, the $1/Q^2$ asymptotic behavior
of $K(0,Q^2)$ has   nothing to do with the
fact that the  contribution  of each particular
bound state has $ 1/Q^2$ behavior.
If,   instead of $\Phi(z)$, one would take a  function
with $\sim z^{\Delta}$ behavior for small $z$,
one would still be able to write the GVMD
representation for such a version of $K(0,Q^2)$,
but its asymptotic behavior will be $\sim 1/Q^{2+\Delta}$.

\subsection{Two deeply virtual  photons}

Each term of the
bound-state  decomposition (\ref{Ank})   for $K(Q^2,Q^2)$
has $1/Q^4$   behavior.  Thus, in  the  case of  strong dominance
of  a few lowest states, one  would expect $1/Q^4$
large-$Q^2$   behavior of  this function.

However, as we already obtained, the function $K(Q^2,Q^2)$
behaves like $1/Q^2$   for large $Q^2$.
This  result was a  consequence of two
features of the form factor integral (\ref{kQQ}).
The   first is the fact that the bulk-to-boundary propagator ${\cal J} (Q,z)$
behaves like $e^{-Qz}$ for   large $Q$.
This    is  a very  general property:
in this limit  ${\cal J} (Q,z)$ should coincide with
its free-field version ${\cal K} (Qz) =  zQ K_1 (zQ)$.
The second  feature is   that the pion wave
function $\Phi (z)$ is   finite at the origin,
which  follows from the basic formula
(\ref{piondecay2})  that defines $f_\pi$.

Hence, to qualitatively understand the mechanism
that produces $1/Q^2$ result from the doubly-infinite sum
of $1/Q^4$  terms,  one may  consider 
a  simpler ``toy''   model that  also  has these general 
properties, but  allows to analytically   calculate integrals 
that determine the coefficients   in
Eq.~(\ref{Ank}).  
In particular, an explicit result  for $K(Q_1^2,Q_2^2)$
may  be obtained if we take the soft-model  expression \cite{Grigoryan:2007my}
  \begin{align}
{\cal J}^{\rm s} (Q,z) =  a \int_0^{1} x^{a-1} \, \exp \left [ -\frac{x}{1-x} \,  \kappa^2 z^2 \right ] \, dx
 \end{align}
for   the bulk-to-boundary propagators (with
$\kappa$ being the oscillator parameter and
$a= Q^2/4 \kappa^2$)
and
  \begin{align}
   \Psi^{\rm s} (z) =e^{-\kappa^2z^2}
  \end{align}
for the pion wave function.
This  model has the required properties,  namely,
 ${\cal J}^{\rm s} (Q,z) $ approaches the free-field
function ${\cal K}(Qz)$  for large $Q^2$, while $\Phi (0)$ is  finite.

Calculating  the integral
  \begin{align}
   K^{\rm s}(Q_1^2,Q_2^2) = 2\kappa^2 \int_0^\infty {\cal J}^{\rm s} (Q_1,z) \, {\cal J}^{\rm s} (Q_2,z) \,
e^{-\kappa^2z^2}  \, z  dz
  \end{align}
gives
\begin{widetext}
  \begin{align}
\label{Ksoft}
   K^{\rm s}(Q_1^2,Q_2^2) =&  \sum_{n=0}^\infty \frac{a_1}{(a_1+n)(a_1+n+1)}
 \frac{a_2 }{(a_2+n)(a_2+n+1)} \ ,
  \end{align}
where $a_1= Q_1^2/M^2$, $a_2= Q_2^2/M^2$ and $M=2 \kappa$
is   the mass of the lowest bound state.
The spectrum corresponding to  ${\cal J}^{\rm s} (Q,z) $
is given by $M_m^2 = mM^2$, with $m=1,2, \ldots$ , and
Eq.~(\ref{Ksoft})
explicitly displays the bound state  poles.
For large $Q_1^2,Q_2^2$,
each  term of this  sum behaves  like $1/Q_1^2 Q_2^2$.
However, taking $a_1=a_2=a \gg 1$ gives
 \begin{align}
   K^{\rm s}(Q^2,Q^2) \to a^2 \int_0^\infty \frac{dn}{(n+a)^4} = \frac1{3a} = \frac{M^2}{3Q^2} \  .
 \end{align}
Thus,   the conversion from the $1/Q^4$ asymptotics of individual
terms to the $1/Q^2$   asymptotics of the sum
is  due to nondecreasing ${\cal   O}(n^0)$  behavior of the coefficients
accompanying $n^{\rm th}$  term of the sum.
In other words, transitions  involving higher bound states
are important, i.e., the  pole decomposition
is   far from   being dominated by  a  few lowest
states.

\subsection{Structure of two-channel pole decomposition}

However, to make specific statements
about the transitions, one should
realize that   Eq.~(\ref{Ksoft})   does not have the form
of Eq.~(\ref{Ank}). In particular, it is  not a  double sum,
and having a   sum over a single
parameter implies that the summation
parameters $n$ and $k$ in the double sum
representation are correlated.
A simple inspection of Eq.~(\ref{Ksoft})  shows that
either $k=n$ or $k=n \pm 1$.
Furthermore, the representation (\ref{Ksoft})
has $Q_1^2 Q_2^2$  factor in the numerator,
which should be canceled against the denominator factors
to get an expression in which $Q_1^2 $ and $Q_2^2$
appear in denominators only.
The easiest way to  obtain the desired
GVMD-type expansion for $K^{\rm s}(Q_1^2,Q_2^2)$
is  to use another \mbox{representation \cite{Grigoryan:2007my}}
  \begin{align}
{\cal J}^{\rm s} (Q,z) =   \kappa^2 z^2
 \int_0^{1} \exp \left [ -\frac{x}{1-x} \,  \kappa^2 z^2 \right ] \, \frac{ x^{a} \,  dx }{(1-x)^2}
 \end{align}
for   the bulk-to-boundary propagators.  Then

  \begin{align}
\label{Ksoft2}
   K^{\rm s}(Q_1^2,Q_2^2) =  \sum_{n=0}^\infty \frac{(n+1)(n+2)}{(a_1+n+1)(a_1+n+2)
(a_2+n+1)(a_2+n+2)} \ .
  \end{align}
Now, each term of the sum decreases as $1/Q_1^4 Q_2^4$, but the sum may be rewritten as
  \begin{align}
\label{Ksoft3}
   K^{\rm s}(Q_1^2,Q_2^2) =
 \sum_{n=1}^\infty  \frac1{1+Q_1^2/M_n^2}  \biggl \{ \frac{2 }{1+Q_2^2/M_n^2}
 -  \frac{1}{1+Q_2^2/M_{n+1}^2} -  \frac{1}{1+Q_2^2/M_{n-1}^2} \biggr \} \, ,
  \end{align}
i.e.,  the coefficients $B^{\rm s}_{n,k}$ of the bound-state expansion (\ref{Ank}) in this   model are  given  by
  \begin{align}
   B^{\rm s}_{n,k} =  2 \delta_{n,k} - \delta_{n,k+1} - \delta_{n,k-1} 
  \end{align}
(there is  no  need   to add  surface terms in this  model since $\Psi^{\rm s} (\infty)=0$, and 
hence  $A^{\rm s}_{n,k}=0$).

\end{widetext}

When $Q_2^2 =0$,   only the $n=1$ term   contributes,
and  $K (Q^2,0)$  in this model is formally  dominated by the lowest   resonance:
  \begin{align}
    K^{\rm s}(Q^2,0) = \frac1{1+Q^2/M_1^2}  \  .
  \end{align}
This fact may create an impression
that  the $\pi^0 \to \gamma \gamma $
decay in this model is  dominated
by the $\rho \omega $ intermediate state.
However, the  outcome that the sums of 
 bracketed terms  are zero for $n\geq 2$  is  a result of cancellation
of  the contribution of a  diagonal transition
that gives 2, and two off-diagonal
transitions,  each of which gives $-1$.

In fact, the coefficients  $B^{\rm s}_{n,n}=2$ of diagonal transitions
do not depend on  $n$, and their
total contribution into  $K^{\rm s}(Q_1^2,Q_2^2)$ diverges. On the other  hand, the coefficients  $B^{\rm
s}_{n,n+1}=-1$ and $B^{\rm s}_{n+1,n}=-1$ 
of subdiagonal  transitions are negative, and
also do not depend on $n$. 
 The  total contribution into  $K^{\rm s}(Q_1^2,Q_2^2)$  of each of $k=n+1$
or $k=n-1$ off-diagonal  transitions also  diverges. In   such a situation,  claiming the dominance of the lowest
states contribution  makes no sense.

In the hard-wall  model, the diagonal coefficients $B_{n,n}$ decrease with $n$,  
and they 
are visibly  larger than the neighboring off-diagonal ones  (see Table I).
Thus,  one may   say that, 
for small $Q_1^2, Q_2^2$, the value of 
$K(Q_1^2,Q_2^2)$  is  dominated by  the lowest bound states
(the   coefficients $A_{n,n}$ also decrease with $n$, see Table II, asymptotically 
they   behave 
like $2\Psi (z_0)/n$). 
\begin{table}
\label{table1}
 \begin{tabular}{| c | c | c | c | c | c |}
\hline
 $k  \diagdown n $ & 1 & 2  & 3  & 4  &5  \\
\hline
1  & 0.7905  & 0.0272  & -0.1331 & 0.0562  &  -0.0275  \\ \hline
2  & 0.0272 &0.2484  &-0.0423  &-0.0608  & 0.0287 \\ \hline
3  & -0.1331 & -0.0423  & 0.1624 & -0.0367 & -0.0403 \\  \hline
4 & 0.0562 & -0.0608 & -0.0367 &0.1199  & -0.0303  \\  \hline
5 &-0.0275  & 0.0287 & -0.0403 &-0.0303  & 0.0951 \\ \hline
\end{tabular}
\caption{Coefficients $B_{n,k}$ in the hard-wall  model.}
\end{table}
\begin{table}
\label{table2}
 \begin{tabular}{| c | c | c | c | c | c |}
\hline $k  \diagdown  n $ & 1 & 2  & 3  & 4  &5  \\
\hline 1  & 0.3641 & -0.2420 &  0.1935 & -0.1658 & 0.1474  \\
\hline 2  & -0.2420 & 0.1609 & -0.1286 & 0.1102 & -0.0980  \\
\hline 3  & 0.1935 & -0.1286 & 0.1028 & -0.0881 &0.0783  \\
\hline 4  & -0.1658 &0.1102 & -0.0881 &0.0755 & -0.0671  \\
\hline 5  &0.1474 & -0.0980 &  0.0783 & -0.0671 &0.0597 \\ \hline
\end{tabular}

\caption{Coefficients $A_{n,k}$ in the hard-wall  model}
\end{table}
However,  the  large-$Q_1^2, Q_2^2$
expansion is   given by
 \begin{align}
  K(Q_1^2,Q_2^2) &=  \frac{M_1^4}{Q_1^2 Q_2^2} \sum_{n,k=1}^\infty 
\frac{M_n^2 M_k^2}{M_1^4} \, (A _{n,k} + B_{n,k}) + \ldots 
\  ,
 \end{align}
i.e., one should deal 
with the coefficients 
 \begin{align}
B_{n,k}\frac{M_n^2 M_k^2}{M_1^4 } = 
B_{n,k}\frac{ \gamma_{0,n}^2 \gamma_{0,k}^2}{\gamma_{0,1}^4 } \  \  {\rm and }
\ \  A_{n,k}\frac{M_n^2 M_k^2}{M_1^4 } \ ,
 \end{align}
the lowest of which are given in Tables III and IV.  Again, the coefficients
increase with $n$  and $k$  producing divergent series,
just like in the toy soft-like model.   Note, that  asymptotically the sum of 
$A$-type terms  gives a  contribution exponentially decreasing 
with $Q_1$ and/or $Q_2$, i.e., much faster than the $\sim 1/Q^4$ 
asymptotic behavior of each particular transition.   
On the other hand,   the sum of 
$B$-type terms  gives a  contribution that has $\sim 1/Q^2$ 
asymptotic behavior, i.e., it drops  slower than $1/Q^4$.

\begin{table}
\label{table3}
\begin{tabular}{| c | c | c | c | c | c |}
\hline $k \diagdown n $ & 1 & 2  & 3  & 4  &5  \\
\hline 1  & 0.7905 & 0.1433 & -1.7236 & 1.3514 & -1.0593 \\
\hline 2  & 0.1433 & 6.8952 & -2.8875 & -7.6956 & 5.8234 \\
\hline 3  & -1.7236 & -2.8875 & 27.2323 & -11.4306 & -20.1136 \\
\hline 4 & 1.3514 & -7.6956 & -11.4306 & 69.3299  & -28.0684  \\
\hline 5 & -1.0593 & 5.8234 & -20.1136 & -28.0684 & 141.2499 \\ \hline
\end{tabular}
\caption{Coefficients $B_{n,k}M^2_n M^2_k/M^4_1 $ in the hard-wall  model.}
\end{table}

\begin{table}
\label{table4}
 \begin{tabular}{| c | c | c | c | c | c |}
\hline $k  \diagdown  n $ & 1 & 2  & 3  & 4  &5  \\
\hline 1  & 0.3641 & -1.2751 & 2.5056 & -3.9868 & 5.6816  \\
\hline 2  & -1.2751 & 4.4655 & -8.775 & 13.9624 & -19.8979  \\
\hline 3  & 2.5056 & -8.775 & 17.2437 & -27.4374 & 39.1011  \\
\hline 4  & -3.9868 & 13.9624 & -27.4374 & 43.6572 & -62.216  \\
\hline 5  & 5.6816 & -19.8979 & 39.1011 & -62.216 & 88.6642 \\ \hline
\end{tabular}

\caption{Coefficients $A_{n,k}M^2_n M^2_k/M^4_1 $ in the hard-wall  model}
\end{table}

The convergence situation may be different  
in the real-world QCD, in which higher resonances are broad, with the width increasing
with $n$ (or $k$). Then the  diagonal and  neighboring non-diagonal transitions strongly overlap
for large $n$  and may essentially cancel
each other. 

\section{Summary}

At the end of the  pioneering paper \cite{Erlich:2005qh},  it was indicated that 
one of the future developments of the holographic
models would be an  incorporation of the 5D Chern-Simons term to reproduce the chiral anomaly of QCD. 
However,   only
relatively recently  Ref.~\cite{Domokos:2007kt} discussed a  holographic model of QCD that 
includes Chern-Simons term (see also \cite{Katz:2007tf}) and,  furthermore,  extends  the 
gauged $ SU(2)_L \otimes SU(2)_R $ flavor group to $ U(2)_L
\otimes U(2)_R $. 

In the present paper, we develop  an  extension of the AdS/QCD model, similar in form to that proposed in
\cite{Domokos:2007kt},  but adjusted  to study the anomalous coupling  of 
the neutral pion to two (in general,
virtual) photons. The additional   part of the  gauge field in the 5D bulk is associated with
the isoscalar vector current (related to $\omega$-like mesons). The Chern-Simons term allows to reproduce the
tensor structure of the anomalous  form factor $F_{\gamma^* \gamma^*
\pi^0}(Q_1^2,Q_2^2)$.  To exactly reproduce the QCD anomaly result for real photons,
we  added   contributions  localized at the IR boundary $z=z_0$,
and then    studied the  momentum dependence of the $F_{\gamma^* \gamma^*
\pi^0}(Q_1^2,Q_2^2)$  form factor in 
  our model.   

In particular, we calculated the slope of the form factor with one real
and one slightly off-shell photon. Our result $a_\pi \approx 0.031$ for the 
parameter of the usual $F_{\gamma^* \gamma^*
\pi^0}(0,Q^2) = F_{\gamma^* \gamma^*
\pi^0}(0,0) (1-a_\pi Q^2/m_\pi^2)$  low-$Q^2$ experimental representation 
of the data is very  close to the value 
$a_\pi = 0.0326\pm 0.0026$  obtained by CELLO collaboration~\cite{Behrend:1990sr} 
from spacelike $Q^2$ measurements, and rather close to the central values $a_\pi \sim  0.024$ 
of two most recent  experiments
\cite{Farzanpay:1992pz,MeijerDrees:1992qb} for timelike  $Q^2$.

Although the holographic model is expected to work for low energies, where QCD is in the strong coupling 
regime, we found it interesting to investigate the behavior   
 of the model form factor also in the  regions where at least one 
of the photon virtualities is 
 large.  For the case with one real and one highly virtual photon,  we demonstrated that our  AdS/QCD
result  is in full agreement with the magnitude of the leading-order  part of the next-to-leading-order 
pQCD fits of existing
experimental data.  In the kinematics  when  both photons have equal and large
virtualities we obtained  the same result as in the leading-order pQCD. Finally, we 
considered the 
general case of unequal and large photon virtualities. In this  case, the form factor has a nontrivial
dependence on the ratio $Q_1^2/Q_2^2$  of photon  virtualities. Our calculation shows  that 
 the final result of our  AdS/QCD model 
 {\it analytically}  coincides  with the pQCD expression calculated using 
  the asymptotic distribution amplitude
$ \varphi_\pi^{{\rm as}}(x) = 6 x (1-x) $.  This  result is rather unexpected,
because initial  expressions   for the   form factor have very different structure. 
It should be noted   that the absolute normalization of the 
form factor  $K(Q_1^2, Q_2^2)  $ 
in our model is  fixed by adjusting its value to $K(0,0)=1$ at the real  photon point,
which allows to conform to the QCD axial anomaly.
The outcome that this choice exactly reproduces 
the leading-order pQCD result for the equal-virtualities form factor $K(Q^2, Q^2)  $
needs further studies, as well as our 
result that the $\omega$-dependence of the 
unequal-virtualities form factor $K((1+\omega)Q^2,(1-\omega)Q^2)$
coincides  with  the leading-order pQCD result 
derived by assuming  the asymptotic shape for the pion distribution amplitude.

The bulk-to-boundary propagators entering into  AdS/QCD formulas  for   form factors 
have a generalized vector-meson-dominance (GVMD) decomposition.
As a result, the   form factors also can  be written in GVMD form.
We   studied the interplay   between the GVMD decomposition of    form factors 
and their behavior for large photon virtualities.
In the case of one real photon, the   function $K(0,Q^2)$ asymptotically behaves like $1/Q^2$.   
However, we demonstrated that this    behavior 
has   nothing to do with the fact that each  term of the GVMD expansion for
$K(0,Q^2)$ also behaves like $1/Q^2$   for large $Q^2$.  In fact, a  formal  GVMD 
expression for the coefficient of the $1/Q^2$   term diverges. 
When   both  photons are  highly virtual,  each  term of the GVMD expansion for $K(Q^2,Q^2)$
behaves like $1/Q^4$, while $K(Q^2,Q^2)$ has $1/Q^2$    asymptotic  behavior. 
Thus, we observe that only in the region of small photon   virtualities it makes sense 
to talk about  dominating role of the lowest states. In particular, for real photons, when $Q_1^2=Q_2^2=0$, 
the   lowest (``$\rho\, \omega \,\pi$'') transition amplitude contributes 
$1.15$  into the $K(0,0)=1$  value, the excess being primarily cancelled by the neighboring 
non-diagonal transitions. 


\vspace{7mm}

\section{Acknowledgments}

H.G. would like to thank C.~D.~Roberts and J.~Goity for 
valuable comments, T.~S.~Lee, M.~Vanderhaeghen, J.~Erlich
and J.~Harvey for  stimulating discussions,
 J.~P.~Draayer and A.~W.~Thomas for support at Louisiana State
University and Jefferson Laboratory.

This paper is authored by Jefferson Science Associates,
LLC under U.S. DOE Contract No. DE-AC05-06OR23177. The U.S.
Government retains a non-exclusive, paid-up,
irrevocable, world-wide license to publish or reproduce this
manuscript for U.S. Government purposes.



\begin{references}


\bibitem{anomaly} 
  S.~L.~Adler,
  Phys.\ Rev.\  {\bf 177}, 2426 (1969);
  J.~S.~Bell and R.~Jackiw,
  Nuovo Cim.\  A {\bf 60}, 47 (1969); 
  J.~S.~Schwinger,
  Phys.\ Rev.\  {\bf 82}, 664 (1951).

\bibitem{cornwall} J.M. Cornwall, {Phys.Rev.Lett.} {\bf 16}, 1174  (1966).

\bibitem{Gross:1972dd}
  D.~J.~Gross and S.~B.~Treiman,
  Phys.\ Rev.\  D {\bf 4}, 2105 (1971).

\bibitem{Brodsky:1971ud}
  S.~J.~Brodsky, T.~Kinoshita and H.~Terazawa,
  Phys.\ Rev.\  D {\bf 4}, 1532 (1971).

\bibitem{Parisi:1971tf}
  J.~Parisi and P.~Kessler,
  Lett.\ Nuovo Cim.\  {\bf 2}, 755 (1971).

\bibitem{Jacob:1989pw}
  M.~Jacob and T.~T.~Wu,
  Phys.\ Lett.\  B {\bf 232}, 529 (1989).




%
\bibitem{Lepage:1980fj}
  G.~P.~Lepage and S.~J.~Brodsky,
  Phys.\ Rev.\  D {\bf 22}, 2157 (1980).
%
\bibitem{Brodsky:1981rp}
  S.~J.~Brodsky and G.~P.~Lepage,
  Phys.\ Rev.\  D {\bf 24}, 1808 (1981).
%
\bibitem{Lepage:1982gd}
  G.~P.~Lepage, S.~J.~Brodsky, T.~Huang and P.~B.~Mackenzie,
  {\it ``Hadronic Wave Functions In QCD''},
Report CLNS-82-522 (1982), published in Proceedings of
1981 Banff Summer Inst.  (1982)
%
\bibitem{Chernyak:1983ej}
  V.~L.~Chernyak and A.~R.~Zhitnitsky,
  Phys.\ Rept.\  {\bf 112}, 173 (1984).
%
\bibitem{Efremov:1979qk}
  A.~V.~Efremov and A.~V.~Radyushkin,
  Phys.\ Lett.\  B {\bf 94}, 245 (1980).
%
\bibitem{Radyushkin:1977gp}
  A.~V.~Radyushkin,
JINR-P2-10717 (1977); English translation: 
  arXiv:hep-ph/0410276.
%
\bibitem{Lepage:1979zb}
  G.~P.~Lepage and S.~J.~Brodsky,
  Phys.\ Lett.\  B {\bf 87}, 359 (1979).
%
\bibitem{Behrend:1990sr}
  H.~J.~Behrend {\it et al.}  [CELLO Collaboration],
  Z.\ Phys.\  C {\bf 49}, 401 (1991).
%
\bibitem{Gronberg:1997fj}
  J.~Gronberg {\it et al.}  [CLEO Collaboration],
  Phys.\ Rev.\  D {\bf 57}, 33 (1998)
%
\bibitem{delAguila:1981nk}
  F.~del Aguila and M.~K.~Chase,
  Nucl.\ Phys.\  B {\bf 193}, 517 (1981).
%
\bibitem{Braaten:1982yp}
  E.~Braaten,
  Phys.\ Rev.\  D {\bf 28}, 524 (1983).
%
\bibitem{Kadantseva:1985kb}
  E.~P.~Kadantseva, S.~V.~Mikhailov and A.~V.~Radyushkin,
  Yad.\ Fiz.\  {\bf 44}, 507 (1986)
  [Sov.\ J.\ Nucl.\ Phys.\  {\bf 44}, 326 (1986)].
%
\bibitem{Musatov:1997pu}
  I.~V.~Musatov and A.~V.~Radyushkin,
  Phys.\ Rev.\  D {\bf 56}, 2713 (1997)
%
\bibitem{Khodjamirian:1997tk}
  A.~Khodjamirian,
  Eur.\ Phys.\ J.\  C {\bf 6}, 477 (1999)
%
\bibitem{Schmedding:1999ap}
  A.~Schmedding and O.~I.~Yakovlev,
  Phys.\ Rev.\  D {\bf 62}, 116002 (2000)
%
\bibitem{Bakulev:2007jv}
  A.~P.~Bakulev, S.~V.~Mikhailov, A.~V.~Pimikov and N.~G.~Stefanis,
  arXiv:0710.2275 [hep-ph].
%
\bibitem{Voloshin:1982ea}
  M.~B.~Voloshin,
Preprint ITEP-8-1982.

\bibitem{Nesterenko:1982dn}
  V.~A.~Nesterenko and A.~V.~Radyushkin,
  Sov.\ J.\ Nucl.\ Phys.\  {\bf 38}, 284 (1983)  
  [Yad.\ Fiz.\  {\bf 38}, 476 (1983)].


\bibitem{Mikhailov:1988nz}
  S.~V.~Mikhailov and A.~V.~Radyushkin,
  Sov.\ J.\ Nucl.\ Phys.\  {\bf 49}, 494 (1989)
  [Yad.\ Fiz.\  {\bf 49}, 794 (1988)].
%
\bibitem{Ito:1991pv}
  H.~Ito, W.~W.~Buck and F.~Gross,
  Phys.\ Rev.\  C {\bf 45}, 1918 (1992).
%
\bibitem{Radyushkin:1995pj}
  A.~V.~Radyushkin,
  Acta Phys.\ Polon.\  B {\bf 26}, 2067 (1995)
%
\bibitem{Anikin:1995cf}
  I.~V.~Anikin, M.~A.~Ivanov, N.~B.~Kulimanova and V.~E.~Lyubovitskij,
  Z.\ Phys.\  C {\bf 65}, 681 (1995).
%
\bibitem{Radyushkin:1996tb}
  A.~V.~Radyushkin and R.~Ruskov,
  Nucl.\ Phys.\  B {\bf 481}, 625 (1996)
%
\bibitem{Kroll:1996jx}
  P.~Kroll and M.~Raulfs,
  Phys.\ Lett.\  B {\bf 387}, 848 (1996);
  R.~Jakob, P.~Kroll and M.~Raulfs,
  J.\ Phys.\ G {\bf 22}, 45 (1996)
%
\bibitem{Anisovich:1996hh}
  V.~V.~Anisovich, D.~I.~Melikhov and V.~A.~Nikonov,
  Phys.\ Rev.\  D {\bf 55}, 2918 (1997)
%
\bibitem{Kekez:1998rw}
  D.~Kekez and D.~Klabucar,
  Phys.\ Lett.\  B {\bf 457}, 359 (1999)
%
\bibitem{Anikin:2000rq}
  I.~V.~Anikin, A.~E.~Dorokhov and L.~Tomio,
  Phys.\ Part.\ Nucl.\  {\bf 31}, 509 (2000)
  [Fiz.\ Elem.\ Chast.\ Atom.\ Yadra {\bf 31}, 1023 (2000)].
%
\bibitem{Maris:2002mz}
  P.~Maris and P.~C.~Tandy,
  Phys.\ Rev.\  C {\bf 65}, 045211 (2002)
%
\bibitem{Dorokhov:2002tq}
  A.~E.~Dorokhov, M.~K.~Volkov and V.~L.~Yudichev,
  Phys.\ Atom.\ Nucl.\  {\bf 66}, 941 (2003)
  [Yad.\ Fiz.\  {\bf 66}, 973 (2003)]
%
\bibitem{Xiao:2005af}
  B.~W.~Xiao and B.~Q.~Ma,
  Phys.\ Rev.\  D {\bf 71}, 014034 (2005)
%
\bibitem{Ruiz Arriola:2006ii}
  E.~Ruiz Arriola and W.~Broniowski,
  Phys.\ Rev.\  D {\bf 74}, 034008 (2006)
%
\bibitem{Maldacena:1997re}
  J.~M.~Maldacena,
  Adv.\ Theor.\ Math.\ Phys.\  {\bf 2}, 231 (1998)
  [Int.\ J.\ Theor.\ Phys.\  {\bf 38}, 1113 (1999)];
  S.~S.~Gubser, I.~R.~Klebanov and A.~M.~Polyakov,
  Phys.\ Lett.\ B {\bf 428}, 105 (1998);
  E.~Witten,
  Adv.\ Theor.\ Math.\ Phys.\  {\bf 2}, 253 (1998)
%
\bibitem{Polchinski:2002jw}
  J.~Polchinski and M.~J.~Strassler,
  Phys.\ Rev.\ Lett.\  {\bf 88}, 031601 (2002);
  JHEP {\bf 0305}, 012 (2003)
%
\bibitem{Boschi-Filho:2002vd}
  H.~Boschi-Filho and N.~R.~F.~Braga,
  JHEP {\bf 0305}, 009 (2003);
  Eur.\ Phys.\ J.\  C {\bf 32}, 529 (2004)
%
\bibitem{Brodsky:2003px}
  S.~J.~Brodsky and G.~F.~de T\'eramond,
  Phys.\ Lett.\ B {\bf 582}, 211 (2004);
%
  G.~F.~de Teramond and S.~J.~Brodsky,
  Phys.\ Rev.\ Lett.\  {\bf 94}, 201601 (2005)
%
\bibitem{Sakai:2004cn}
  T.~Sakai and S.~Sugimoto,
  Prog.\ Theor.\ Phys.\  {\bf 113}, 843 (2005);
%
   {\bf 114}, 1083 (2006)
%
\bibitem{Erlich:2005qh}
  J.~Erlich, E.~Katz, D.~T.~Son and M.~A.~Stephanov,
  Phys.\ Rev.\ Lett.\  {\bf 95}, 261602 (2005)
%
\bibitem{Erlich:2006hq}
  J.~Erlich, G.~D.~Kribs and I.~Low,
  Phys.\ Rev.\ D {\bf 73}, 096001 (2006)
%
\bibitem{DaRold:2005zs}
  L.~Da Rold and A.~Pomarol,
  Nucl.\ Phys.\ B {\bf 721}, 79 (2005);
%
  JHEP {\bf 0601}, 157 (2006)
%
\bibitem{Karch:2006pv}
  A.~Karch, E.~Katz, D.~T.~Son and M.~A.~Stephanov,
  Phys.\ Rev.\ D {\bf 74}, 015005 (2006)
%
\bibitem{Csaki:2006ji}
  C.~Csaki and M.~Reece,
  JHEP {\bf 0705}, 062 (2007)

\bibitem{Hambye:2005up}
  T.~Hambye, B.~Hassanain, J.~March-Russell and M.~Schvellinger,
  Phys.\ Rev.\  D {\bf 74}, 026003 (2006);
{\it ibid.} 
{\bf 76}, 125017 (2007).
%
\bibitem{Hirn:2005nr}
  J.~Hirn and V.~Sanz,
  JHEP {\bf 0512}, 030 (2005);
%
  J.~Hirn, N.~Rius and V.~Sanz,
  Phys.\ Rev.\ D {\bf 73}, 085005 (2006)
%
\bibitem{Ghoroku:2005vt}
  K.~Ghoroku, N.~Maru, M.~Tachibana and M.~Yahiro,
  Phys.\ Lett.\  B {\bf 633}, 602 (2006)
%
\bibitem{Brodsky:2006uq}
  S.~J.~Brodsky and G.~F.~de Teramond,
  Phys.\ Rev.\ Lett.\  {\bf 96}, 201601 (2006)
%
\bibitem{Evans:2006dj}
  N.~Evans, A.~Tedder and T.~Waterson,
  JHEP {\bf 0701}, 058 (2007)
%
\bibitem{Casero:2007ae}
  R.~Casero, E.~Kiritsis and A.~Paredes,
  Nucl.\ Phys.\  B {\bf 787}, 98 (2007).
%
\bibitem{Gursoy:2007cb}
  U.~Gursoy and E.~Kiritsis,
  JHEP {\bf 0802}, 032 (2008).
%
      \bibitem{Gursoy:2007er}
        U.~Gursoy, E.~Kiritsis and F.~Nitti,
        JHEP {\bf 0802}, 019 (2008).


%
\bibitem{Bergman:2007pm}
  O.~Bergman, S.~Seki and J.~Sonnenschein,
  JHEP {\bf 0712}, 037 (2007).
%
%
\bibitem{Erdmenger:2007vj}
  J.~Erdmenger, K.~Ghoroku and I.~Kirsch,
  JHEP {\bf 0709}, 111 (2007).
%
%
\bibitem{Erdmenger:2007cm}
  J.~Erdmenger, N.~Evans, I.~Kirsch and E.~Threlfall,
  arXiv:0711.4467 [hep-th].
%


\bibitem{Dhar:2007bz}
  A.~Dhar and P.~Nag,
  JHEP {\bf 0801}, 055 (2008).





\bibitem{Grigoryan:2007vg}
  H.~R.~Grigoryan and A.~V.~Radyushkin,
  Phys.\ Lett.\  B {\bf 650}, 421 (2007).
%
\bibitem{Grigoryan:2007wn}
  H.~R.~Grigoryan and A.~V.~Radyushkin,
  Phys.\ Rev.\  D {\bf 76}, 115007 (2007).
%



\bibitem{Grigoryan:2007iy}
  H.~R.~Grigoryan,
  Phys.\ Lett.\  B {\bf 662}, 158 (2008).





\bibitem{Abidin:2008ku}
  Z.~Abidin and C.~E.~Carlson,
  arXiv:0801.3839 [hep-ph].
%



\bibitem{Kwee:2007dd}
  H.~J.~Kwee and R.~F.~Lebed,
  JHEP {\bf 0801}, 027 (2008).



\bibitem{Witten:1998qj}
  E.~Witten,
  Adv.\ Theor.\ Math.\ Phys.\  {\bf 2}, 253 (1998).
%


%
\bibitem{Hill:2006wu}
C.~T.~Hill,
Phys.\ Rev.\  D {\bf 73}, 126009 (2006).

\bibitem{Domokos:2007kt}
  S.~K.~Domokos and J.~A.~Harvey,
  Phys.\ Rev.\ Lett.\  {\bf 99}, 141602 (2007).
%

%
\bibitem{Chern:1974ft}
  S.~S.~Chern and J.~Simons,
  Annals Math.\  {\bf 99}, 48 (1974).

%
\bibitem{Grigoryan:2007my}
  H.~R.~Grigoryan and A.~V.~Radyushkin,
  Phys.\ Rev.\  D {\bf 76}, 095007 (2007).
%
\bibitem{Fujiwara:1984mp}
  T.~Fujiwara, T.~Kugo, H.~Terao, S.~Uehara and K.~Yamawaki,
  Prog.\ Theor.\ Phys.\  {\bf 73}, 926 (1985).


\bibitem{Wess:1971yu}
  J.~Wess and B.~Zumino,
  Phys.\ Lett.\  B {\bf 37}, 95 (1971).


\bibitem{Witten:1983tw}
  E.~Witten,
  Nucl.\ Phys.\  B {\bf 223}, 422 (1983).


%



%
\bibitem{Meissner:1987ge}
  U.~G.~Meissner,
  Phys.\ Rept.\  {\bf 161}, 213 (1988).
%




 
%
\bibitem{Farzanpay:1992pz}
  F.~Farzanpay {\it et al.},
  Phys.\ Lett.\  B {\bf 278}, 413 (1992).
\bibitem{MeijerDrees:1992qb}
  R.~Meijer Drees {\it et al.}  [SINDRUM-I Collaboration],
  Phys.\ Rev.\  D {\bf 45}, 1439 (1992).
%
\bibitem{Fonvieille:1989kj}
  H.~Fonvieille {\it et al.},
  Phys.\ Lett.\  B {\bf 233}, 65 (1989).
%
\bibitem{primex}
 A. Gasparian, et al. ``A Precision Measurement of the Neutral
  Pion Lifetime via the Primakoff Effect'', JLab proposal PR-02-103,
  http://www.jlab.org/primex/
%


\bibitem{Nesterenko:1982gc}
  V.~A.~Nesterenko and A.~V.~Radyushkin,
  Phys.\ Lett.\  B {\bf 115}, 410 (1982).


\bibitem{Katz:2007tf}
  E.~Katz and M.~D.~Schwartz,
  JHEP {\bf 0708}, 077 (2007)
  [arXiv:0705.0534 [hep-ph]].
%



\end{references}
\end{document}